\documentclass{elsart}

\usepackage{amssymb}

\usepackage{amsmath}
\usepackage{graphicx}
\usepackage{epsfig}

\newcommand{\be}{\begin{equation}}
\newcommand{\ee}{\end{equation}}
\newcommand{\bma}{\begin{displaymath}}
\newcommand{\ema}{\end{displaymath}}

\begin{document}
\begin{frontmatter}

\title{Metal clusters, quantum dots and trapped atoms -- from single-particle
  models to correlation}

\author[matti]{M. Manninen} and
\author[steffi]{S.M.Reimann}

\address[matti]{Nanoscience Center, Department of Physics,
FIN-40014 University of Jyv\"askyl\"a, Finland}

\address[steffi]{Mathematical Physics, Lund Institute of Technology,
SE-22100 Lund, Sweden}

\date{today}

\begin{abstract} 
In this review, we discuss the electronic structure of finite quantal systems 
on the nanoscale. 
After a few general remarks on the many-particle physics of the harmonic
oscillator -- likely being the most studied example for the many-body systems
of finite quantal systems, we turn to the electronic 
structure of metal clusters. We discuss Jahn-Teller deformations for the 
so-called 'ultimate' jellium model which assumes a complete cancellation 
of the electronic charge with the ionic background.
Within this model, we are also able to understand the stable electronic shell 
structure of tetrahedral (3D) or triangular (2D) cluster geometries,
resembling closed shells of the harmonic confinement, but for Mg clusters 
being ``doubly-magic'' as the electronic shells occur at precisely
twice the atom numbers in the close-packed tetrahedra. 
Taking a turn to the physics of quantum dot artifical atoms, we discuss the
electronic shell structure of the quasi two-dimensional, harmonically confined 
electron gas. Between the clear shell closings, corresponding to the 
magic numbers in 2D, Hund's rule acts, maximizing the quantum dot spin 
at mid-shell. 
After a brief excursion to multicomponent  quantum dots and the formation 
of Wigner molecules, we turn to finite quantal systems in strong magnetic
fields or, equivalently, electron droplets that are set highly rotating. 
Working within the lowest Landau level, we draw the analogy between 
magnetic fields and rotation, commenting on the formation of the so-called
maximum density droplet (MDD) and its edge reconstruction beyond the integer 
quantum Hall regime. 
Formation and localization of vortices beyond the MDD, as well as electron
localization at extreme angular momenta, are discussed in detail. 
Analogies to the bosonic case, and the systematic build-up of the 
vortex lattice of a rotating Bose-Einstein condensate at high angular momenta, 
are drawn. 
With our contribution we wish to emphasize the 
many analogies that exist between metallic clusters, 
semiconductor artificial atoms, and cold atoms in traps.   
\end{abstract}
\end{frontmatter}

\section{Introduction}

One common feature of most small quantum-mechanical systems is the
discreteness of the quantum states. In systems with high symmetry
the single-particle energy levels are degenerate, which may lead
to  shell structure. This is known to happen in free atoms, but also
in nuclei~\cite{BM}. Spherical metal clusters~\cite{deheer1993}, 
where the particles move in a spherically symmetric mean field, 
provide another example.
In semiconductor quantum dots with circular 
symmetry, shell structure was observed by conductance 
spectroscopy by Tarucha {\it et al.}~\cite{tarucha1996}. 
(For a review, see~\cite{reimann2002}). 
The 'universality' of shell structure bridges these fields 
of physics. However, there are also fundamental differences:
In atoms and in quantum dots, the fixed external 
potential dominates, leading to Hund's rule with maximum spin at mid-shell 
to resolve the degeneracy of the spherical confinement. 
The valence electrons in metal clusters, or the neutrons and
protons of a nucleus, however, move in a mean-field potential determined solely
by the particle dynamics. To resolve degenaracies for non-closed shells, 
metal clusters and nuclei exhibit spontaneous shape deformation,  
while atoms and quantum dots do not. 
Consequently, the often used name 'artificial atom' is well
suited for semiconductor quantum dots, but would be misleading for
free metal clusters.

Many properties of metal clusters can be calculated by using so-called
ab initio electronic structure calculations and molecular
dynamics. These computational results often are in very good agreement with 
experimental data --  
as, for example, in photoemission spectroscopy.
They can pin-point 
the detailed ground-state geometries of  particular
cluster sizes~\cite{akola1999}. 
However, many overall features can even be 
understood using simple models~\cite{deheer1993,brack1993}. 
This also holds for
semiconductor quantum dots, where often, simple single-particle 
models have been very successful~\cite{reimann2002}.  

The purpose of this (brief, and by no means complete) 
review is to summarize the simple models, their
advantages and limitations in describing overall properties of metal
clusters and quantum dots, and to draw analogies between these 
finite quantal systems to the more recently emerging field of cold (bosonic or
fermionic) atom gases in traps. 

Let us begin by looking at the simple, but relevant
many-particle physics of the harmonic oscillator. These results are then
applied to understand the jellium model for metal clusters and
electronic states in quantum dots. The universality of deformation
is shortly described using simple models.  
We finally turn to a comparsion of quantum dots at strong magnetic fields, 
and weakly interacting bosonic systems that are set rotating.

\section{Many-particle physics in harmonic oscillator}
\label{harmonic}

The harmonic oscillator confining a single-particle is solved in about all text
books of quantum mechanics. However, adding more particles 
immediately makes it more challenging to describe the system theoretically, 
and new interesting phenomena appear.
The many-body Hamiltonian is then written as 
\be
H=\sum_i
\left(-\frac{\hbar^2}{2m}\nabla_i^2+\frac{1}{2}m\omega_0^2r_i^2\right)
+\sum_{i\ne j}v(\vert{\bf r}_i-{\bf r}_j\vert),
\label{h1}
\ee
where $\omega_0$ is the frequency of the confining harmonic oscillator
and $v(r)$ the interparticle two-body interaction. 
The position vector $r_i$ and the
Laplace operator $\nabla_i^2$ may be three-, two- or one-dimensional
depending on the system in question.
Sometimes we want to use the occupation number representation and 
write
\be 
H=\sum_{i,\sigma} \epsilon_{i} c_{i,\sigma}^+ c_{i,\sigma}
+\sum_{\{i,\sigma\}} v_{i_1,i_2,i_3,i_4} c_{i_1,\sigma_1}^+
c_{i_2,\sigma_2}^+ c_{i_4,\sigma_4} c_{i_3,\sigma_3},
\label{h2}
\ee
where $c^+$ and $c$ are the normal creation and annihilation
operators (as here, for fermions),
and $\epsilon_i$ is the single-particle energy of the form
$\epsilon_i=\hbar\omega_0(n_i+d/2)$, $d$ being the dimension of the
system.
Most conveniently, one  uses the single-particle states of the 
confining harmonic oscillator as a basis. 
It is important to note that even if the spin index appears in
this formulation of the Hamiltonian (as a summation index), 
here we consider only spinless interactions, i.e. the Hamiltonian is, 
as obvious from Eq.~(\ref{h1}), independent of the spin.

The perhaps most important feature of a harmonic confinement is, that 
the center-of-mass motion separates from the internal motion,
regardless of the interaction between the particles.
This can easily be shown for both classical and 
quantum systems~\cite{talmi1993}.
As a consequence, the selection rule for the dipole oscillations
only allows the center-of-mass excitation. 
In the case of simple metal clusters and quantum dots this is 
the plasmon resonance, with energy $\hbar\omega_0$, where 
$\omega_0$ is the frequency of the harmonic
confinement~\cite{brack1993}.
In connection with two-dimensional quantum dots~\cite{pfannkuche1991,wixforth1994,hawrylak1995,darnhofer1995,heitmann1997}, the effect of
the separation of the center-of-mass motion was earlier often referred to
as ``Kohn's theorem''~\cite{kohn1961}.
In the case of atomic nuclei, the related excitation is called 
the ``giant resonance'', where the proton and neutron distributions 
oscillate with respect to each other~\cite{BM}.

Another important property of the harmonic confinement is 
the separation of the spatial coordinates from the center-of-mass
 motion (or single-particle motion). This means that the level
structure in the most general case (with different oscillation frequencies
$\omega_i$ along different directions) is simply
\be
\epsilon_{n_1,n_2,n_3}=\hbar\omega_1(n_1+\frac{1}{2})
+\hbar\omega_2(n_2+\frac{1}{2})+\hbar\omega_3(n_3+\frac{1}{2}).
\ee
For spherically symmetric potentials, labelling the energy levels 
by their radial and angular momentum indices,
one obtains the harmonic energy shells given in Table~1.
Including the spin degree of freedom with a factor of two, the 
``magic numbers'' of the harmonic oscillator in three dimensions
occur  at particle numbers 
$2, 8, 20, 40, 70,...$, and at $2, 6, 12, 20, 30, 42,...$
in two dimensions.
\begin{table}[h]
\caption{Shell structure of three and two-dimensional 
(3D and 2D) harmonic oscillators. $g$ is the degeneracy of the shell 
and $N$ the cumulatice number of states without spin degeneracy.}
\label{hoshells}
\begin{tabular}{l|l|ccc|lccc}
\hline
 & & 3D & & & & 2D & & \\
shell&levels& g & $N$ & $2N$&~~~g& $N$ & $2N$\\
\hline
1&$1s$~~~&1&1&2  &~~~1&1&2\\
2&$1p$&3&4&8        &~~~2&3&6\\
3&$2s1d$&6&10&20     &~~~3&6&12\\
4&$2p1f$&10&20&40    &~~~4&10&20\\
5&$3s2d1g$&15&35&70  &~~~5&15&30\\
6&$3p2f1h$&21&56&112 &~~~6&21&42\\
\hline
\end{tabular}
\end{table}
In a non-harmonic potential, the additional degeneracy of different 
radial states disappears, and other degeneracies occur.
A famous example is the Wood-Saxon potential, 
\be
V_{\rm WS}(r)=-\frac{V_0}{1+e^{(R-r)/a}}~,
\label{ws}
\ee
frequently used in nuclear physics 
(where the spin-orbit interaction of the nucleons further 
splits the shells~\cite{BM}). 
Single-particle states in this potential are filled following 
the sequence $1s, 1p, 1d, 2s, 2p, 1f,$ etc.~. 
The Woods-Saxon potential Eq.~(\ref{ws})
is a good approximation for the mean-field
potential in metal clusters~\cite{frauendorf1992}, 
with energetically dominant magic numbers at 
2, 8, 20, 40, 58, 92, etc. Note, however, that the first few magic numbers, 
as here $2, 8$ and 20,  
are mainly  determined by the angular momentum degeneracy and are nearly 
independent of the radial shape of the spherical potential.

In the 2D harmonic confinement the noninteracting electron states can
be solved analytically also in the presence of 
a magnetic field~\cite{reimann2002}, 
the resulting single-particle states being the so-called Fock-Darwin 
states~\cite{fock1928,darwin1930,landau1930}. 
Consequently, the harmonic confinement has been widely 
utilized when studying the quantum Hall effect 
in finite systems~\cite{laughlin1983}.

\section{Jellium model of metal clusters}

Many properties of simple metals, like alkalis, alkali earths and even
aluminum can be explained as properties of the interacting, homogeneous
electron gas. The role of the ions is then merely to keep the electron
gas together at its equilibrium density. In the jellium
model~\cite{lang1973}, inside the metal the charge 
of the ions is smoothened out and replaced by a homogeneous background charge
with the same density as the electron gas.
At the surface, the background charge goes abruptly to zero.
The electron density
is usually described by the density parameter $r_s$ defined as the 
radius of a sphere (in units of Bohr radius) containing one electron:
$4\pi r_s^3/3=1/n_0$, with $n_0$ being the number density of the electrons.
The density functional method in the local density approximation (LDA)
is ideally suited for the jellium model which naturally has a smooth
and slowly varying electron density.

This approach was 
first used to study metal surfaces~\cite{lang1970}, 
lattice defects~\cite{manninen1975}
and impurities in metals~\cite{puska1981}.
The first application to metal clusters was made by 
Martins {\it et  al.}~\cite{martins1981}, who studied the size variation 
of the ionization energy. Similar work had been successful for calculating 
the work 
function of planar surfaces of alkali metals~\cite{monnier1978}.

In the density functional Kohn-Sham method 
the electrons more in an effective mean-field potential
\be
V_{\rm eff}^\sigma({\bf r})=-e\phi({\bf r})+V_{\bf xc}^\sigma({\bf r}),
\ee
where $\phi$ is the total electrostatic potential of the background
charge and electron density distribution, and $V_{\rm xc}$
is the exchange-correlation potential which depends locally on the 
electron density and spin polarization~\cite{parr1989}.
In the case of a spherical jellium cluster the effective potential resembles a 
finite potential well with rounded edge. It can be 
well approximated by the above mentioned Woods-Saxon potential (Eq.~\ref{ws}). 
The spherical jellium model suggests that, like in free atoms,
the ionization potential is largest for the magic clusters,
and at a minimum when only one electron occupies an open shell. 
The experimental results, however, show a much richer
structure as a function of the cluster size, which 
in alkali metals is dominated 
by a marked odd-even staggering~\cite{deheer1993}. 
The reason behind are shape deformations, as  described 
in the next section.

In the spherical jellium model for metal clusters, the 
the background charge is a homogeneously charged sphere of
radius $R=\sqrt[3]{N} r_s$. The (external) potential 
caused by this sphere is 
\bma 
V_{\rm sphere}(r)=\left\{ \begin{array}{ll}
-\frac{Ne^2}{8\pi\epsilon_0 R^3} \left(3R^2-r^2\right) & 
\textrm{if $r \le R$} \\
-\frac{Ne^2}{4\pi\epsilon_0 r} & \textrm{if $r>R$} ~.\\
\end{array} \right.
\ema
Note that the potential inside the sphere is harmonic
and can be written as 
\be
V_{\rm sp}(r)=\frac{1}{2}m\omega_{\rm sp}^2r^2,
\ee
where
\be
\omega_{\rm sphere}^2=\frac{e^2}{4\pi\epsilon_0r_s^3m}=
\frac{n_0e^2}{3\epsilon_0m}=\frac{\omega_p^2}{3}
\ee
is the square of the plasmon frequency of a metallic sphere.
Since the electrons only slightly spill out from the region of the 
harmonic potential, the plasmon is the dominating dipole absorption 
mechanism for spherical jellium clusters~\cite{brack1993}.
Ekardt~\cite{ekardt1984} used the spherical jellium model in connection with 
the time-dependent density functional theory to study optical absorption.
He found that the anharmonicity of the background potential
caused fragmentation of the single plasmon peak to a 
distribution of close-lying absorption peaks.
Similar work was subsequently done by several groups, using the RPA 
method~\cite{brack1993,yannouleas1998}. 

Koskinen {\it et al.}~\cite{koskinen1994}
used shell-model methods from nuclear physics to try to
solve the electronic structure and photo-absorption of the 
jellium clusters beyond the mean-field approach. 
For up to eight electrons, they could diagonalize the 
many-electron Hamiltonian nearly exactly. Already
for 20 electrons, however, the configuration interaction method showed
a much too slow convergence as a function of the size of the 
basis set (in fact, it was shown that the error in the 
correlation energy was $\propto E_{\rm cut-off}^{-3/2}$).
For eight electrons, their result agreed with those of the
RPA calculations.
For positively charged jellium spheres, the 
fragmentation of the plasmon peak disappears as the confinement of the 
electrons becomes harmonic~\cite{yannouleas1989,koskinen1994}.

Historically, it is interesting to note that 
Martins {\it et al.}~\cite{martins1981} 
corrected the pure jellium results by including ion pseudo-potentials
via first-order perturbation theory, in a similar fashion than 
Lang and Kohn~\cite{lang1973} had done for metal surfaces.
While for planar surfaces the correction had only a minor effect
(in alkali metals), it became dominating for large metal clusters,
and completely diminished the effects of the electronic shell
structure of the pure jellium sphere. A similarly large effect of
the pseudopotential correction was observed for large
voids in metals, shown to be due to the
low-index surfaces present in spherical systems cut from an ideal 
lattice~\cite{manninen1978}. The notion of the possible importance
of the lattice potential made the theoreticians cautious in making too strong
predictions of the applicability of the jellium model to 
real metal clusters~\cite{hintermann1983,ekardt1984b}, 
until the magic numbers of alkali 
metal clusters were observed~\cite{knight1984}.

The degeneracy of the open shell
clusters should lead to Hund's rules like in the case of free atoms.
In the spherical jellium models the clusters
with open shells should have a large total spin and magnetic 
moment~\cite{hintermann1983}. 
This was predicted prior to the success of the jellium 
model by Geguzin~\cite{geguzin1981}, who studied highly 
symmetric cub-octahedral Na$_{13}$ clusters.
For free clusters, however, deformation wins over Hund's 
rule and removes both the degeneracy and magnetism~\cite{ben_confpaper}.

We conclude that the simple 
spherical shell structure explains well the magic numbers 
in the experimental mass spectra of sodium clusters.
It lies behind the so-called super shells~\cite{balian1970,nishioka1990}
observed in alkali metal clusters~\cite{pedersen1991}, as well as the 
importance of the collective plasmon resonance.

The simple jellium model also accounts for some properties of 
noble metals. Recently, it has been observed that 
even in gold clusters some features can be explained 
most easily with arguments based on the 
jellium model~\cite{walter2006}.


\section{Deformed jellium}

The similarity of small nuclei and 
simple metal clusters is not limited to 
magic numbers and to the existence of the plasmon-type 
giant resonance, but extends even to the internal deformation
of the system. It is clear that the smallest clusters
can be viewed as well-defined molecules with a 
geometry determined by the atomic configurations.
Quantum chemistry can be used to characterize
the ground state and spectroscopic properties of clusters
with only a few atoms. For larger clusters 
($N>10$) the early theories assumed 
spheres cut from a metal lattice~\cite{martins1981b},
or faceted structures with shapes determined by the 
Wulff polyhedra~\cite{zangwill1988}. In reality, however, the
clusters exhibit geometries very different from
these ideal structures. Many metals form
icosahedral clusters~\cite{martin1996,baletto2005}. 
Jahn-Teller deformations are 
important even in quite large clusters, as
manifested, for example, by the odd-even staggering of the 
ionization potential~\cite{deheer1993}.

In the early cluster beam experiments, the temperature
of the clusters was lowered only by evaporative cooling.  
The resulting cluster temperatures were so high that 
the clusters were most likely liquid~\cite{bjornholm1999}. 
The clusters showed the electronic shell structure
as well as deformation, as determined by the 
splitting of the plasmon peak~\cite{chandezon1997}. In fact,
the super-shell structure could only been seen in liquid
sodium clusters. Solid clusters formed
icosahedral structures which governed the abundance and
ionization potential spectra~\cite{martin1994}.

To model  cluster deformations, 
Clemenger~\cite{clemenger1985} was the first to apply the Nilsson
model familiar from nuclear physics~\cite{BM}. 
He was able to explain qualitative
features of the abundance spectrum of sodium clusters,
including the observed odd-even staggering.
A more general model, based on the Strutinsky-model
of nuclei~\cite{brack1972}, 
was developed by Reimann {\it et al.}~\cite{reimann1993}, 
and applied to triaxial geometries by Yannouleas and 
Landmann~\cite{yannouleas1993} as well as 
Reimann {\it et al.}~\cite{reimann1995,reimann1996}.  It could explain 
nearly quantitatively the stabilities and deformation of small
sodium clusters.

\subsection{Ultimate jellium model}

\begin{figure}[h]
\includegraphics[width=0.6\columnwidth]{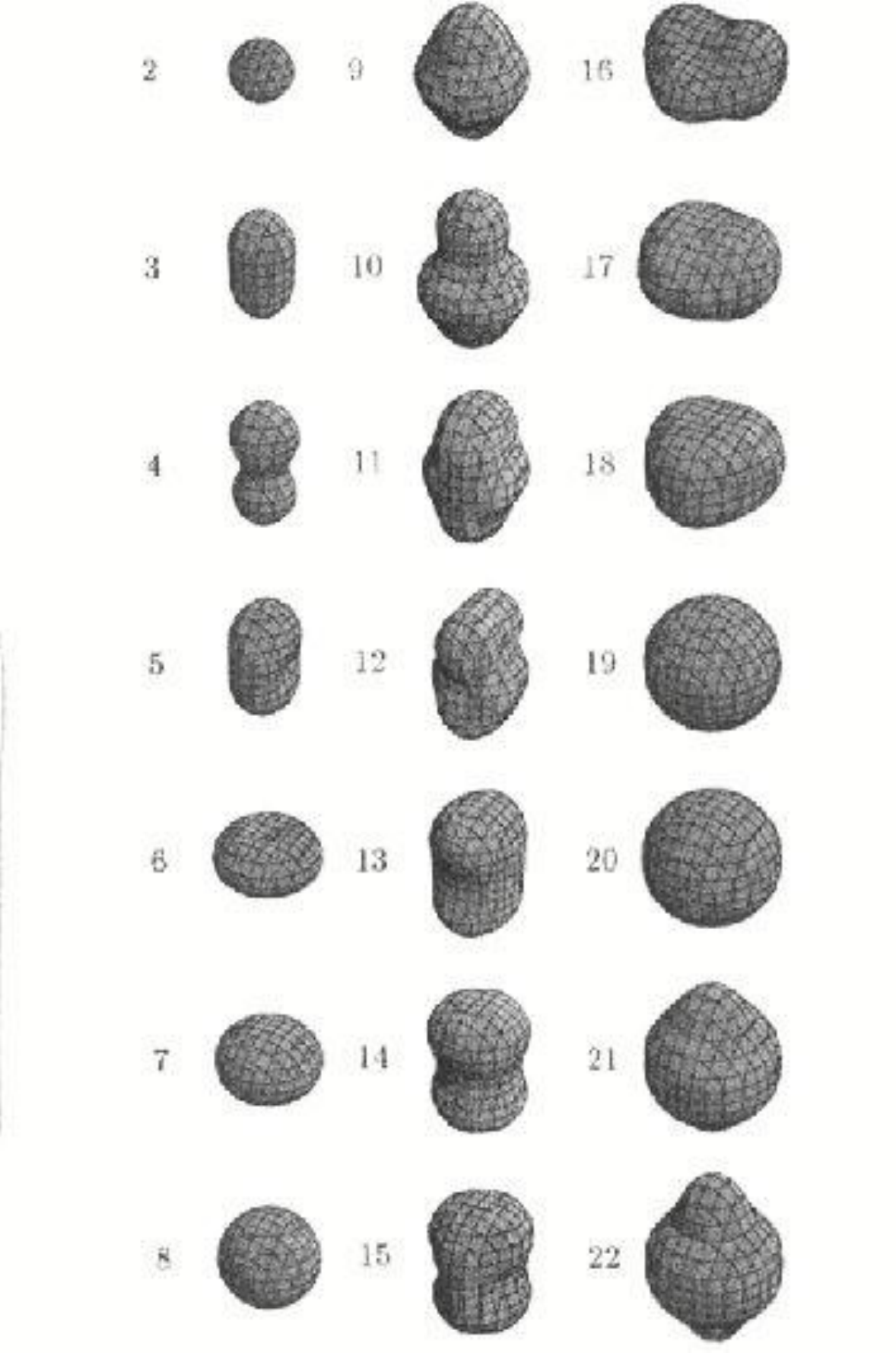}
\caption{Constant-density surfaces of 'ultimate jellium' clusters 
with up to 22 electrons. After Koskinen {\it et al.}, see 
Ref.~\cite{koskinen1995} for details and scales).}
\label{uj3dshapes}
\end{figure}

The simplest way to include deformation in the jellium model
is to assume the uniform background charge density to be a
spheroid, or an ellipsoid~\cite{penzar1990}. The model 
explains qualitatively the splitting of the plasmon peak
and the size dependence of the ionization potential of alkali
metal clusters. However, the optimal deformation shape 
determined by the electronic structure is not an ellipsoid, 
but a more generally shaped jellium background~\cite{hirschmann1994}.
In the ultimate limit, the energy is minimized, when the 
background density equals  the electron density -- as
suggested by Manninen already in 1986~\cite{manninen1986}.
In this so-called 'ultimate jellium model' (UJM)~\cite{koskinen1995},
the density of the background is not fixed,
but in a large cluster adjusts itself to correspond to 
$r_2\approx 4.2~a_0$, a value close to the equilibrium electron 
density in sodium. (Here, $a_0$ is the Bohr radius). 
The ground-state densities for clusters with $N=2$ to 22 electrons 
are shown as constant density surfaces in Fig.~\ref{uj3dshapes}. 
Clearly, the magic numbers at small $N$, here for 2, 8, and 20 electrons, 
correspond to spherical symmetry of the freely deformable 'ultimate jellium' 
droplet. Off-shell, however, the shapes of the clusters exhibit
breaking of axial and inversion symmetries. In general, the resulting 
ground-state geometries are far from ellipsoidal. Clusters which lack
inversion symmetry, are very soft against odd-multipole 
deformations~\cite{koskinen1995}.

\begin{figure}[h]
\includegraphics[width=0.4\columnwidth]{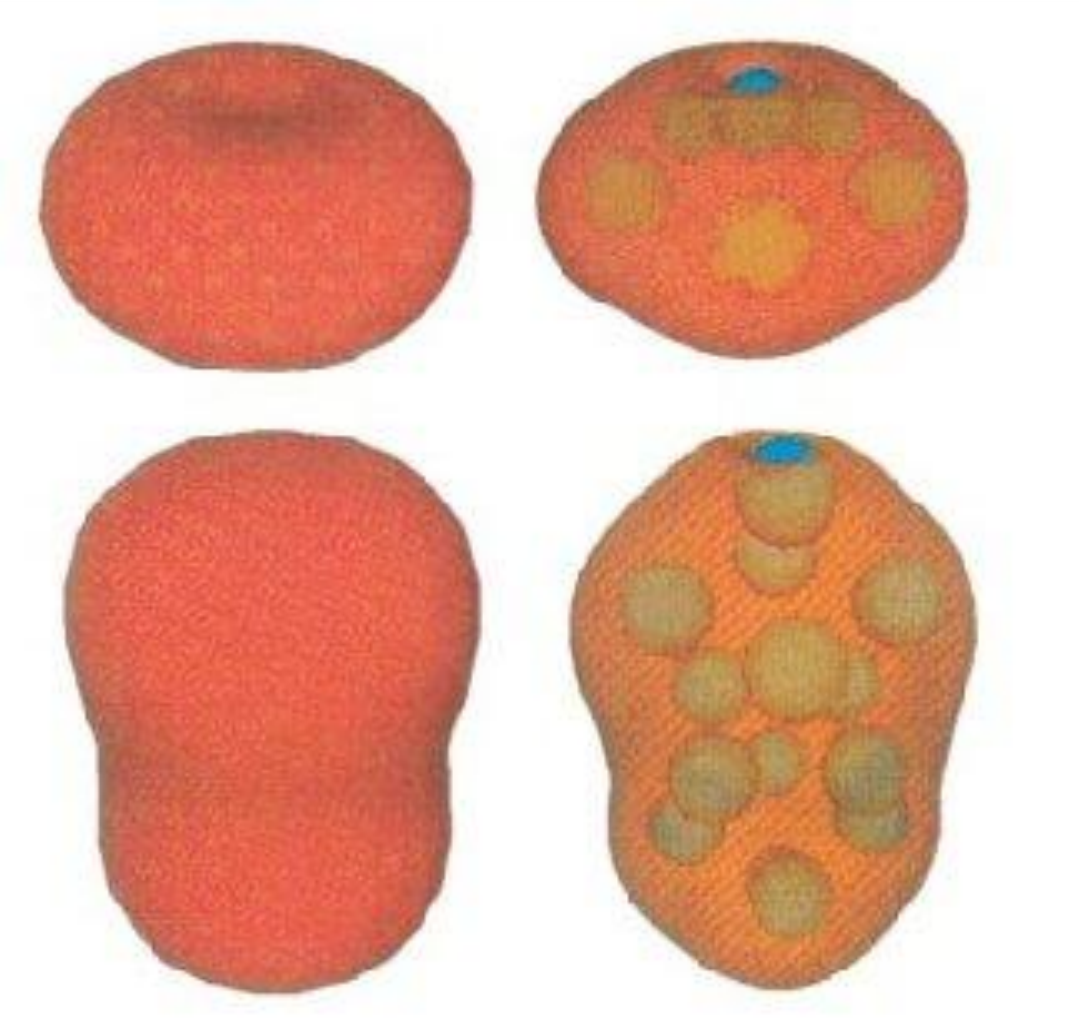}
\caption{Comparison of shapes of UJM clusters (left) to those of 
DFT-LDA molecular-dynamics methods (right), for Na-6 (upper panel)
and Na-14 (lower panel). In all cases the outer surface
shown corresponds to the same particle density. Blue spheres
represent the ions.
From Ref.~\cite{hakkinen1997}.}
\label{hannu_prl}
\end{figure}

\begin{figure}[h]
\includegraphics[width=0.7\columnwidth]{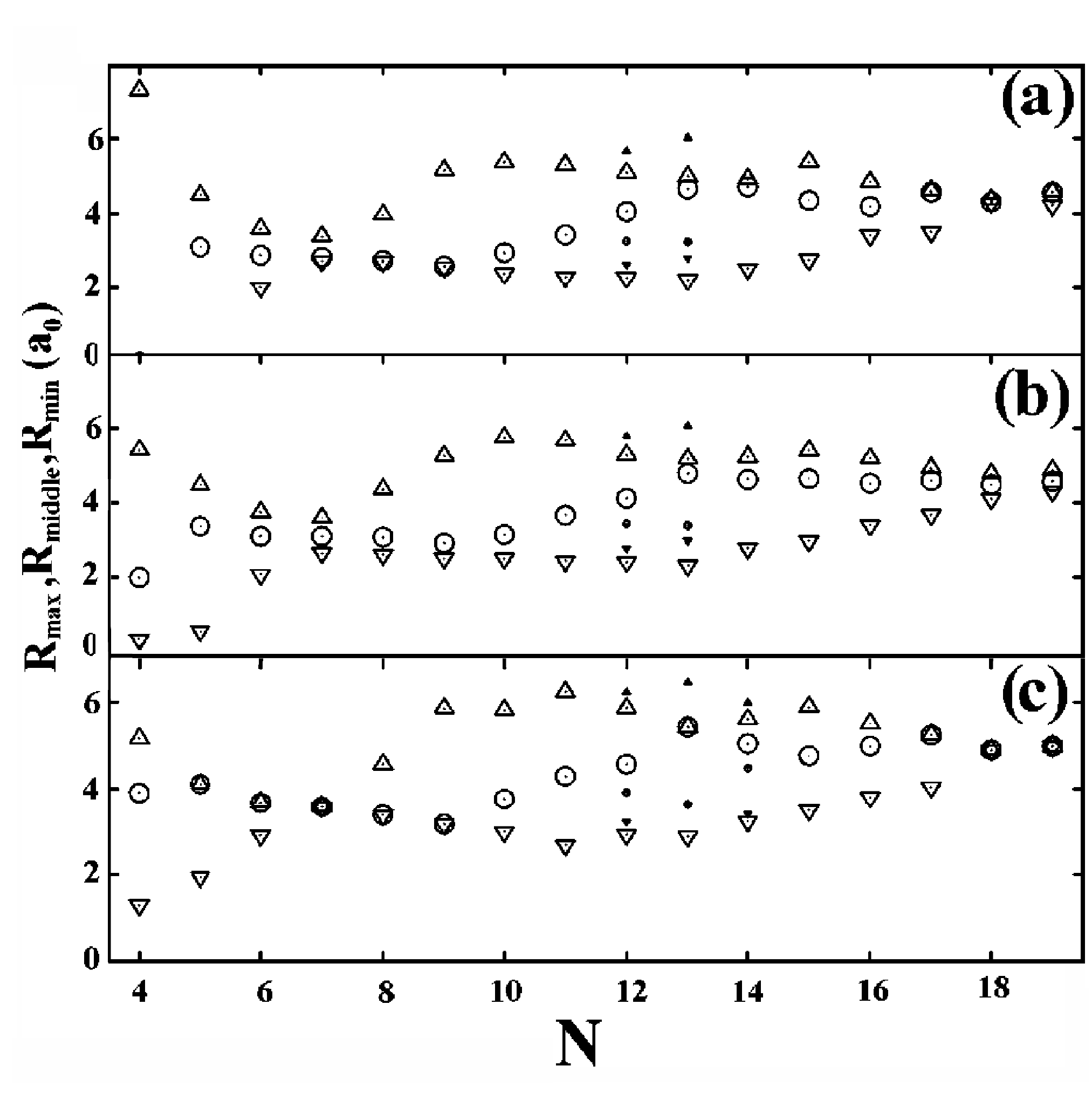}
\caption{The three radii of the anionic
sodium clusters along the principal axis, plotted vs the number of
atoms in the cluster. Down-triangles, circles, and up-triangles
correspond to $R_{\rm min}$ , $R_{\rm middle}$ , and $R_{\rm max}$, 
respectively.
(a) Radii corresponding to the ground-state geometry of {\it ab initio} 
calculations, 
(b) thermally
averaged radii from room-temperature simulations,
(c) radii calculated within the UJM.
From Ref.~\cite{moseler2003}.}
\label{hannu}
\end{figure}

Remarkably, the results obtained from the 
UJM for deformations are very close to
those of ab initio calculations for sodium~\cite{moseler2003},
as shown in Fig.~\ref{hannu_prl} and Fig.~\ref{hannu}. 

Koskinen {\it et al.}~\cite{koskinen1995b} applied the 
UJM to determine the shape deformations of small
nuclei. Their method gave rather 
good agreement with experimental results, and
surprisingly, nearly exactly the same geometries 
as for the electron-gas jellium. 
H\"akkinen {\it et al.}~\cite{hakkinen1997}  studied further the idea 
of this 'universal deformation' and found that in
the LDA, density functional theory predicts similar
deformations for all small fermion clusters.

This shape universality can be easily understood 
in systems where the particles move in a mean field
caused by the particles themselves. When the number of 
the particles is small, there is only a small number
of single-particle states which determine the shape.
For example, for four particles, only the $1s$ and, say, 
$1p_x$ states are filled. Consequently, the shape is
prolate along the $x$-direction. 
This corresponds to the basis of the Nilsson model~\cite{BM}.

The robustness of the shape on the specific model
was further studied by Manninen {\it et al.}~\cite{manninen2000}, 
who showed that deformations of the UJM
are in very good agreement with results of 
the 'ultimate' tight-binding model: the H\"uckel model
for clusters~\cite{yoshida1994}.

\begin{figure}[h]
\begin{tabular}{lr}
\includegraphics[width=0.5\columnwidth]{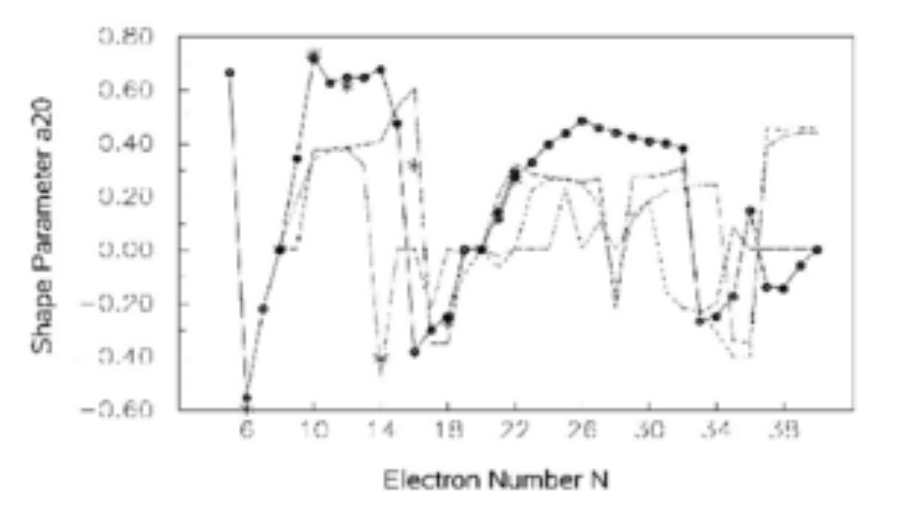} &
\includegraphics[width=0.4\columnwidth]{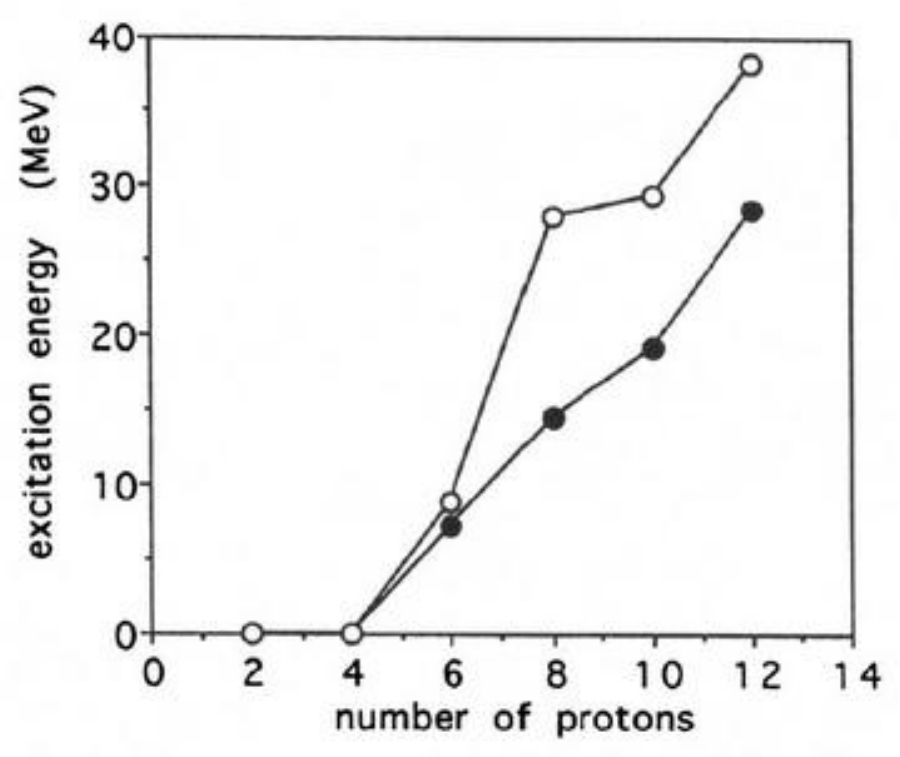}
\end{tabular}
\caption{Left: Shape parameter $a_{20}$ for fermion clusters from 5 to 40
particles calculated with the UJM (black
dots connected with solid line) compared to the experimental
results for even-even nuclei (stars).
Right: Excitation energies of linear isomers calculated with
the UJM for nuclei (open circles) and
compared to the experimental results (black dots). 
From Ref.~\cite{manninen2000}.}
\label{nucldef}
\end{figure} 

For nuclei,  the simple universal
model only needs two parameters, the bulk modulus 
and the average binding energy per nucleon 
(the first term in the so-called mass formula~\cite{BM}),
to give good quantitative approximations to the
deformation parameters and even excitation energies 
of shape isomers, as shown in Fig.~\ref{nucldef}.

\subsection{Triangles and tetrahedra}

\begin{figure}[h]
\includegraphics[width=0.8\columnwidth]{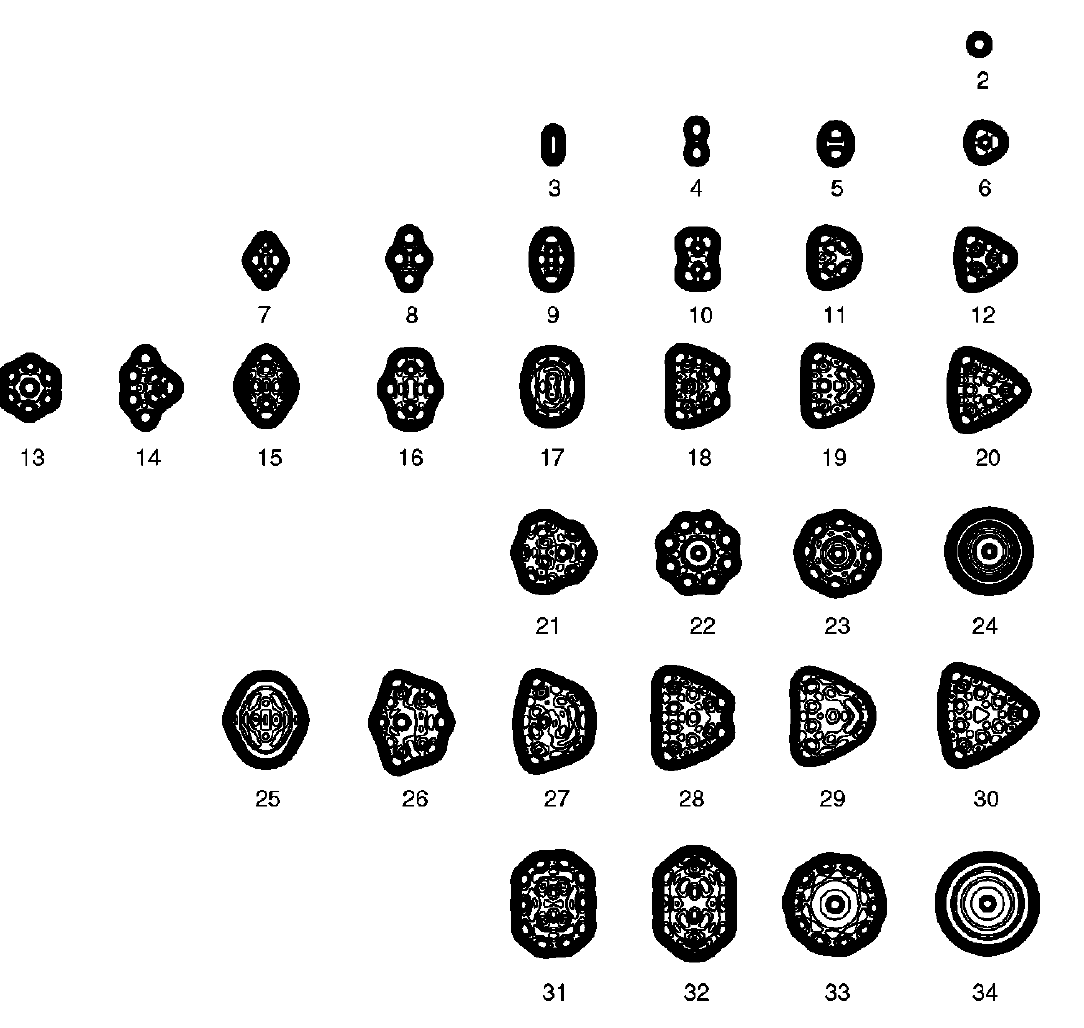}
\caption{Contours of the self-consistent ground-state densities  of
two-dimensional UJM clusters for electron numbers $N\le 2\le 34$, 
calculated for a 2D layer thickness of 3.9$a_0$. From 
Ref.~\cite{reimann1998} (see this Ref. for details).}
\label{2dujmshapes}
\end{figure}

\begin{figure}[h]
\includegraphics[width=0.6\columnwidth ]{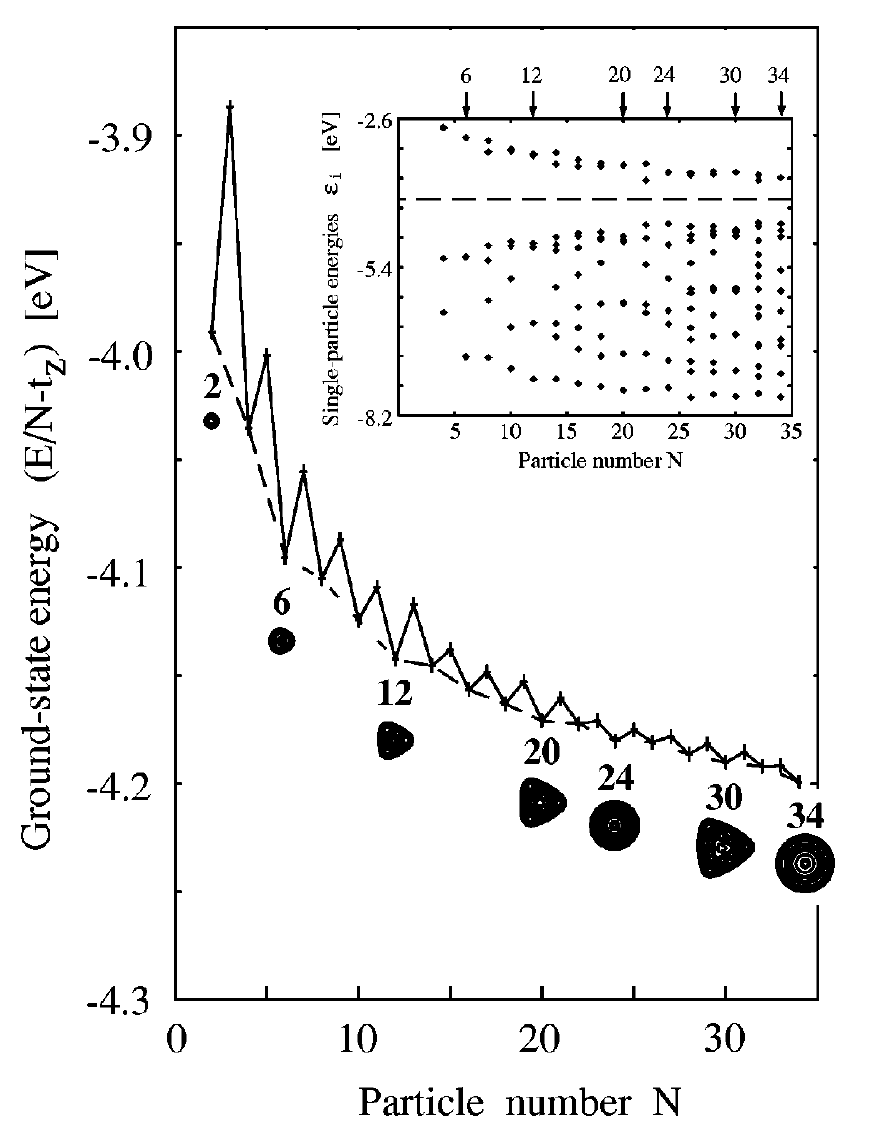}
\caption{Ground-state energies per electron of two-dimensional
clusters, as a function of cluster size 
$N$. (The kinetic energy contribution in $z$-direction, $t_z$, was
  subtracted). The inset shows the self-consistent single-particle Kohn-Sham 
energies for even particle numbers.
From Ref.~\cite{reimann1998}.}
\label{2dujmenergies}
\end{figure}

\begin{figure}[h]
\includegraphics[width=0.7\columnwidth]{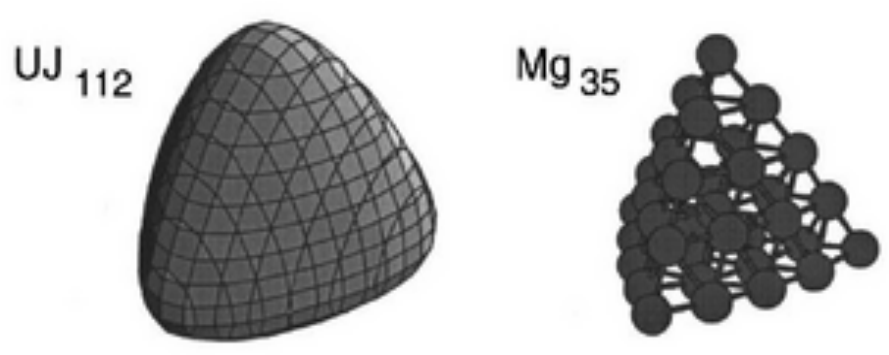}
\caption{Left: Ground-state shape of the UJM for
112 electrons. Right: A possible structure of a Mg$_{35}$ cluster
with 70 valence electrons.
From Ref.~\cite{reimann1998}.}
\label{tetra}
\end{figure}

The jellium model has also been aplied to quasi two-dimensional clusters, 
as for example in the early studies by 
Kohl~{\it et al.}~\cite{kohl1996,kohl1997}. A physical
realization of two-dimensional clusters could be sodium clusters on an inert
surface, or even two-dimensional electron-hole
liquids in semiconductors. 
Reimann {\it et al.}~\cite{reimann1998} analyzed systematically the 
UJM ground-state shapes for quasi two-dimensional sodium clusters.
Contours of the self-consistent ground-state densities  of these 
two-dimensional fermion droplets are shown in Fig.~\ref{2dujmshapes}, 
calculated for a 2D layer  thickness of 3.9$a_0$. 
The shape systematics
reveals that for electron numbers 6, 12, 20, and
30 the 2D clusters have triangular shape. Initially, 
this result appeared puzzling, as these shell closures 
correspond to those of the circular two-dimensional harmonic oscillator, 
and one should thus expect azimuthal symmetries of the ground-state 
densities. The explanation was, however, that in 2D, a triangular 
cavity has precisely these magic numbers~\cite{reimann1997}, and only in the 
large-$N$ limit, the increased surface tension at the corners 
makes the oscillator shells more stable. 
In 2D, the shell closings are rather weak, with favorable energy minima (gaps
at the fermi level) appearing mainly in the small-$N$ limit. 
Given the freedom of unrestricted
shape deformations, a pronounced odd-even staggering appears in the 
ground-state energies, as seen in Fig.~\ref{2dujmenergies}. 
Incidentally, these shell fillings for the triangular geometries 
(without spin-degeneracy) equal precisely the number of atoms forming a 
close-packed triangle. 
In fact, the same holds in three dimensions: at small $N$, 
thetrahedral shell structure 
is prefered~\cite{hamamoto, reimann1998}, 
with magic numbers at $N=2, 8, 20, 40, 70$ and 112. 
These numbers correspond precisely to twice the numbers of atoms in 
a close-packed thetrahedral cluster geometry (see Fig.~\ref{tetra}). 
One should expect that the compact tetrahedral 
geometry at an electronic magic number stabilizes
theses clusters. However, first principles calculations 
have shown that this is not generally the case.
Mg$_{10}$ has an overall tetrahedral shape, but is not a
perfect tetrahedron~\cite{akola2001}.
Na$_{20}$~\cite{rothlisberger1991},  
and Mg$_{20}$~\cite{akola2001} are not tetrahedra,
but Au$_{20}$ seems to be~\cite{li2003}. 
The experimental abundance spectrum of Mg shows 
a maximum at Mg$_{35}$~\cite{diederich2005} --
but so far, there is no
evidence that its geometry is a tetrahedron 
like the one shown in Fig.~\ref{tetra}.
The above results suggests that trivalent metals on
an oxide or graphite surface could favor triangular
shapes.

In fact, advances in the experimental realization of surface-supported planar 
clusters have been recently reported by Chiu {\it et al.}~\cite{chiu2006}.
They found magic numbers in quasi two-dimensional Ag clusters grown on Pb
islands, and studied the transition from electronic to geometric shell 
structure.

We finally mention that high stability of tetrahedral shapes has also been 
discussed in nuclear physics~\cite{dudek2002,dudek2006}, 
predicting tetrahedral 
ground states for some exotic nuclei around $^{110}$Zr 
(see Schunck {\it et al.},~\cite{schunck2004}).

\section{Semiconductor quantum dots}

Generally speaking, a quantum dot is a system where a small number of
electrons are confined in small volume in all three spatial directions. 
It can be, for example, 
a three-dimensional atomic cluster or a two-dimensional island
of electrons formed by external gates in a semiconductor 
heterostructure~\cite{chakraborty1999,reimann2002}. 
In this review, we shall only consider 
{\it two}-dimensional semiconductor quantum dots. Most often they
are formed from AlGaAs-GaAs layered structures, where 
a low-density 2D conduction electron gas is formed in the 
AlGaAs layer. The quantum dot is formed by removing the electrons
outside the dot region with external gates (lateral dot), 
or by etching out the material outside the dot region (vertical 
dot). In both cases, the resulting confining potential is, 
to a good approximation, harmonic. The underlying 
lattice of the semiconductor material 
can be taken into account by using an effective mass
for the conduction electrons, and a static dielectric constant,
reducing the Coulomb repulsion.

The resulting generic model for a semiconductor quantum dot 
is a 2D harmonic oscillator with interacting electrons. 
This in fact is like a 2D jellium model, with the simplification that 
now the harmonic confinement has infinite range and the 
center-of-mass motion separates out exactly (Kohn's theorem~\cite{kohn1961}). 
This means that in the ideal case (in zero magnetic field)
there is only one dipole absorption
peak, as seen in experiments~\cite{demel1990}.

Conductance spectroscopy can be used on one single dot.
The dot is weakly connected to  leads and 
the current is measured as a function of the gate voltage 
which determines the chemical potential and thus the number 
of electrons in the dot~\cite{reimann2002}.
When the electron number in the dot is large, the energy
of an additional electron can be estimated from the 
capacitance $C$ of the dot, as $\Delta E=e^2/C$. The 
resulting conductance then shows equidistant peaks 
as a function of the gate voltage.

\begin{figure}[h]
\includegraphics[width=0.5\columnwidth]{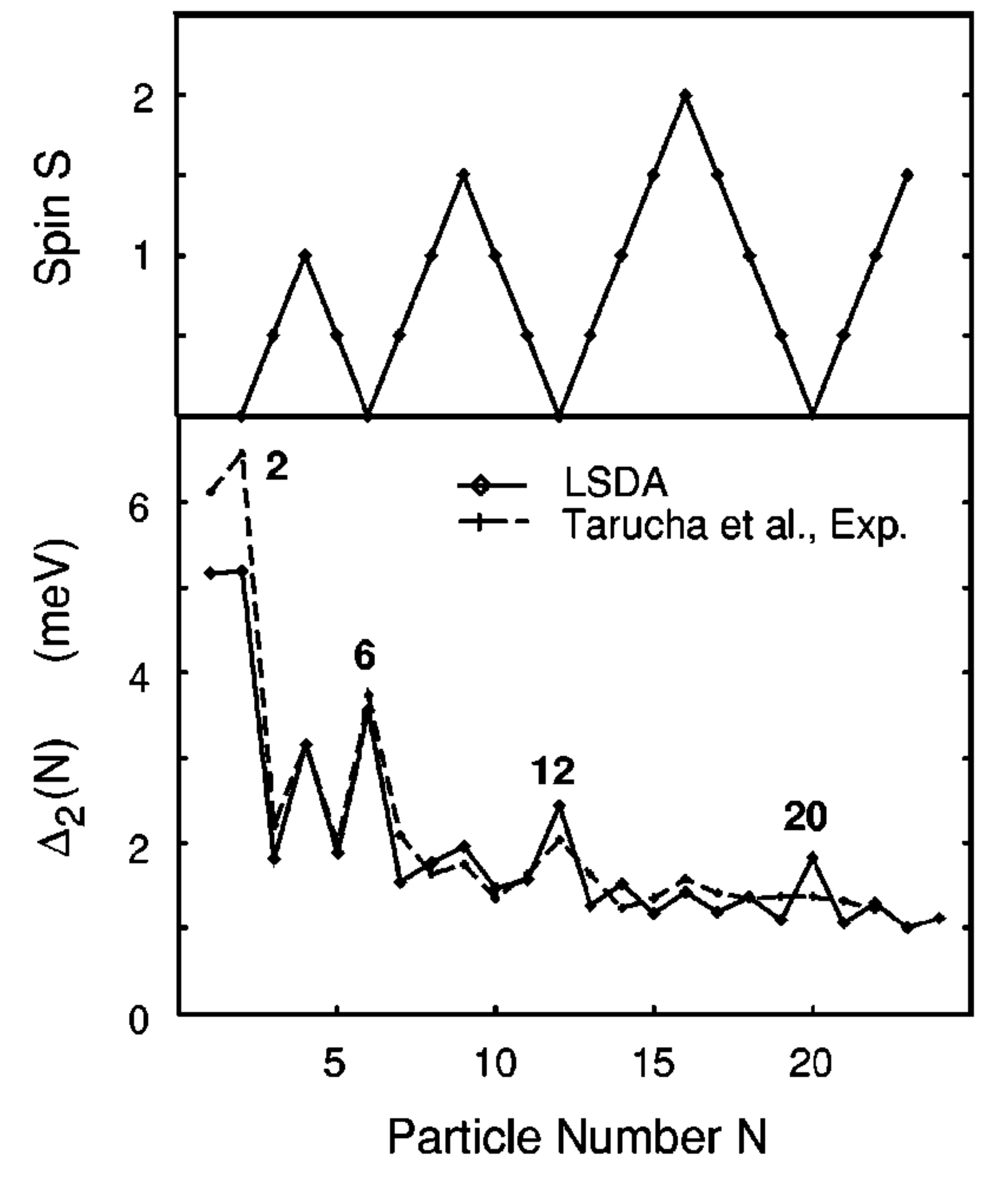}
\caption{Second derivative of the total energy of electrons
in a quantum dot as a function of the number of electrons.
The magic numbers are shown. The experimental result is from
Tarucha {\it et al.}~\cite{tarucha1996}. The upper panel shows the
calculated total spin.}
\label{qdot}
\end{figure}

When the number of electrons is small, the individual
single electron levels in the dot become important and their
shell structure can be seen in the conductance spectrum.
Tarucha {\it et al.}~\cite{tarucha1996} were the first to successfully
determine the shell structure of circular quantum dots.
Their result is shown in the lower panel of Fig.~\ref{qdot}, 
where the 
second derivative of the total energy of the dot is plotted as a function
of the number of electrons, $N$. For comparison, the corresponding 
result of the LSDA calculation for electrons in a harmonic
oscillator is included, too. 

The density functional Kohn-Sham method for semiconductor
quantum dots usually assumes that $i$) the system is two-dimensional,
$ii)$ only the conduction electrons are considered, 
with an effective mass $m^*$ and their Coulomb
interaction screened by the static dielectric function $\epsilon$
of the material in question, and $iii)$ they move in a harmonic
confinement $m\omega_0^2r^2$. A local approximation is used 
for the spin-dependent exchange-correlation energy, derived from the
functionals for the 2D electron gas~\cite{tanatar1989}. For details see
Refs.~\cite{reimann2002}. 

Shell structure with main shell 
fillings (magic numbers) at $N= 2, 6, 12, 20$ appears very clearly in the 
addition energy differences $\Delta _2(N)$. (See Fig.~\ref{qdot}).
Furthermore, like in free atoms, due to Hund's rule at mid-shell 
the total spin is maximal. This means that (just like for the 
spherical jellium model discussed above), any half-filled shell 
shows as a weak 'magic' number, with increased stability. This is clearly 
seen in Fig.~\ref{qdot} where the second derivative
of the total energy shows maxima at $N=4$ and $N=9$ in addition to
the clear peaks at the filled 
shells, $N= 2, 6,$ and  $12$. Figure~\ref{qdot} also 
shows the calculated total spin as a function of the number
of electrons in the dot.
The self-consistent data appear to agree very nicely with the 
experimental data. However, we notice that this agreement becomes worse
with increasing $N$, showing very clear deviations between theory and
experiment after the third shell, i.e. around $N=20$. 
Another series of experimental data, was later published by the same 
group in 2001. In Ref.~\cite{matagne2001}, 
addition energies for 14 different quantum dot structures, all similar 
to the device used in the earlier work by 
Tarucha {\it et al.}~\cite{tarucha1996},  were analyzed. 
Strong variations in the spectra were reported, very clearly differing 
from device to device and seemingly indicating that each of these 
vertical quantum dots indeed has its own properties: a comparison 
to the theoreticalle expected shell structures needs to be taken with care. 
Progress with vertical quantum dots was achieved more recently, 
where few-electron phenomena could be studied by tunneling spectroscopy 
through quantum dots in nanowires~\cite{thelander2003,fuhrer2007}.
 
\begin{figure}[h] 
 \includegraphics[width=0.6\columnwidth]{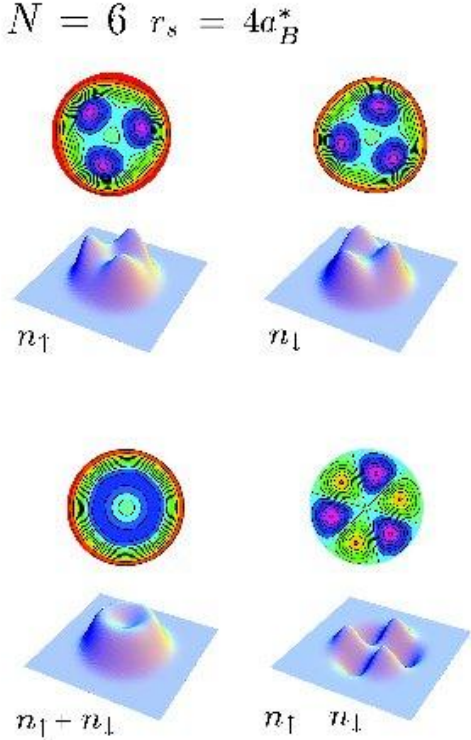}
\caption{DFT spin densities $n^{\uparrow }$ and $n^{\downarrow }$
(upper panel) and total density ($n^{\uparrow }+ n^{\downarrow }$)
as well as (un-normalized) spin polarization ($n^{\uparrow }+ n^{\downarrow
}$) (lower panel) for a six-electron quantum dot at $r_s = 4 a_0^*$,  
shown as 3D plots and their contours. 
From Ref.~\cite{borgh2005}.}
\label{sdw}
\end{figure}

The self-consistent electronic structure calculations 
for quantum dots for some electron numbers showed
internal symmetry-breaking of the spin-density~\cite{koskinen1999},
leading to a static ``spin-density wave'' (SDW). Figure~\ref{sdw}
shows, as an example, the intriguing ground-state spin polarization
for a quantum dot with six electrons. 
For not too small densities of the electron gas, i.e. $r_s\le 6a_0^*$, 
this quantum dot still has a closed-shell configuration, with $S=S_z=0$.
This result is obtained from SDFT.
The total density obtained by the SDFT method is circularly
symmetric, with zero net polarization ($S=0$).  
However, the spin polarization (which equals the difference 
between the spin densities $n^{\uparrow }- n^{\downarrow }$ normalized by the
total density, $n^{\uparrow }+ n^{\downarrow }$), in standard SDFT breaks the 
azimuthal symmetry of the confinement, showing a regular spin 
structure. Fig.~\ref{sdw} shows this very clearly for the example 
of a six-electron quantum dot at $r_s=4 a_0$.
Both spin-up and spin-down densities exhibit three clear bumps, which 
are twisted against each other by an angle of $\pi /3$. This resembles 
very much an antiferromagnet-like structure, with alternating up- and down
spins, on a ring.
Such states were obtained both with the Tanatar-Ceperley~\cite{tanatar1989}
as well as the more recent 
Attaccalite--Moroni--Gori-Giorgi--Bachelet (AMGB) ~\cite{attaccalite2002} 
functionals for 
exchange-correlation. As the AMGB functional depends explicitly on the 
spin polarization, there is left no doubt that the SDW states  
are {\it not} simply 
an artefact of the ad-hoc approximation to the correlation energy,
which is  usually interpolated following the polarization-dependance of 
the exchange energy~\cite{koskinen1999}.
\begin{figure}[h]
\includegraphics[width=0.7\columnwidth]{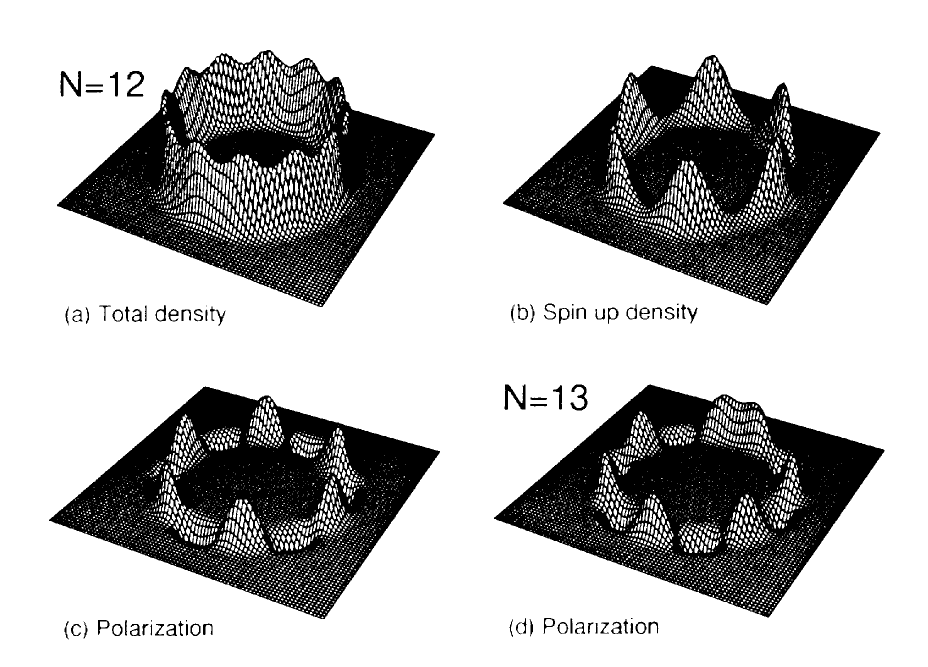}
\caption{Quantum rings with $N=12$ and $N=13$ electrons, showing 
antiferromagnetic spin ordering along the ring. The maximum electron density 
in the 12-electron ring is at $n_{max} = 0.157 {a_0^*}^{-2}$.
From Ref.~\cite{reimann1999}.}
\label{sdwrings}
\end{figure} 
In quasi-one-dimensional quantum rings (see section~\ref{sec:quantrings}), 
these SDW states become more distinctive, as shown in Fig.~\ref{sdwrings}.

The existence of the non-spherical spin-densities in quantum dots 
was disputed in the literature, since the spherical symmetry of the Hamiltonian
dictates spherical symmetry~\cite{hirose1999,harju2004}. 
However, as well known from
nuclear physics, a meal field theory (like KS-LSDA) can lead to
{\it internal} symmetry breaking. In some cases~\cite{borgh2005} 
it reveals the internal structure which, in fact, can be very difficult to
extract from the exact wave function. We will repeatedly meet 
this problem, for example when studying vortices and 
localization in rotating quantum systems, see 
Sections~\ref{sec:vortices1} and \ref{sec:vortices2} below.
(For further reading on the internal symmetry breaking, 
we refer to the recent review articles~\cite{reimann2002,viefers2004,
yannouleas2006}).

The full quantum-mechanical problem of a few 
electrons in a 2D harmonic oscillator
can be solved using the so-called configuration interaction (CI) 
technique, numerically diagonalizing a large Hamiltonian matrix. 
This method is often
called ``exact diagonalization'', although there
is always an approximation due to the necessary restrictions in the 
basis set or the number of configurations included.
Nevertheless, up to say 6 or even 10 particles (depending
on the confinement strength) the results can be viewed as 
practically exact. 

In general, the results of the exact diagonalization agree
well with those obtained within the LSDA. The same spins
dictated by the Hund's rule are obtained and the total
energies agree with good accuracy. Also, the electron-density
and spin-density profiles are in excellent agreement.
However, the exact diagonalization can reveal the existence
of the internal symmetry breaking only via the pair correlation
\be
g_{\sigma\sigma'}({\bf r},{\bf r}')=
\langle \Psi\vert\hat n_\sigma({\bf r})\hat 
n_{\sigma'}({\bf r}')\vert\Psi\rangle
\ee
where $\Psi$ is the many-particle quantum state and $\hat n_\sigma$ the
spin-density operator. This pair correlation function is also
called as ``conditional probability'', since it gives the
probability of finding an electron with spin $\sigma'$ at
${\rm r}'$ when an electron with spin $\sigma$ is located
at ${\bf r}$.
\begin{figure}[h]
\includegraphics[width=0.8\columnwidth]{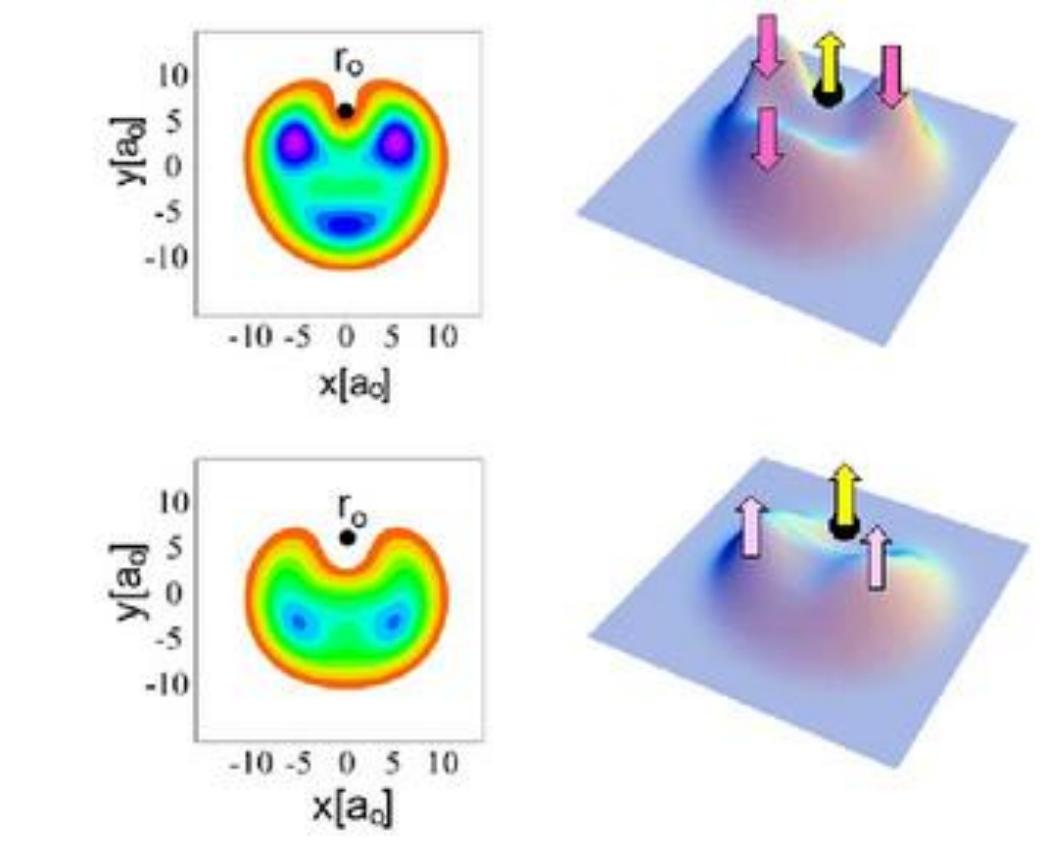}
\caption{Pair correlations 
for a quantum dot with $N=6$ electrons, here at 
$r_s =3.8a_0^*$, obatined with the CI method.  
The top panel shows the (up,down)-correlations, 
the bottom panel the (up-up) correlations. The black dot marks the
reference point, with given spin.  
From Ref.~\cite{borgh2005}.}
\label{exactsdw}
\end{figure} 
As an example, we show in Fig.~\ref{exactsdw} the pair correlations 
for the above discussed six-electron quantum dot, here at 
$r_s =3.8a_0^*$. The top panel shows the (up,down)-correlations, 
the bottom panel the (up-up) correlations. Clearly, the internal structure 
of the exact ground state resembles the SDFT result described above: 
The probability maxima appear on a ring, with six alternating maxima of  
the up- and down correlations.

For a more detailed discussion, we refer to the recent work by 
Borgh {\it et al.}~\cite{borgh2005} on broken-spin-symmetry in SDFT 
ground states and the reliability of SDFT. Here we only note 
that SDFT and CI results generally agree very well in the case of 
{\it singlet} 
states, as it was examplified above by the six-electron quantum dot. 
If the true ground state is a spin-multiplett, however, SDFT introduces 
an artificial splitting of multiplet states which may be misleading, 
and even become a real pitfall when determining ground state energies 
and symmetries where CI or other exact results to compare with, 
are not at hand.

\subsection{Wigner molecules}

The exact diagonalization method has also been used to study 
electron localization in low-density quantum dots. 
At extremely low densities, the homogeneous electron gas forms
a Wigner crystal~\cite{wigner1934} also in the bulk. 
This happens in 3D, 2D and 1D, although in 1D the true long
range order is fading with distance. 

In three dimensions, the critical value at which crystallization 
occurs was determined to be $r_s=100a_0$  
by Ceperley and Alder~\cite{ceperley1980}. 
In two dimensions, the transition occurs at smaller $r_s$ values, 
according to Tanatar and Ceperley~\cite{tanatar1989} at  $r_s >  37{a_0}^*$. 
Breaking of translational invariance in 2D lowers this value to 
$r_s\approx 7.5{a_0}^*$. 
Thus, in finite systems, localization may happen at even smaller values, 
as discussed, for example, by Creffield {\it et al.}~\cite{creffield1999}, 
Egger {\it et al.}~\cite{egger1999}, or Yannouleas and 
Landmann~\cite{yannouleas1999}.

For finite number of electrons in a quantum dot,
the localized state is often called a ``Wigner molecule''~\cite{egger1999}.
The local density approximation can not produce properly the 
localized states due to the lack of exact cancellation of
the direct and exchange Coulomb interactions. 
The most direct notion of electron localization can be
found by using the unrestricted Hartree-Fock
approximation. The (complicated) mean-field 
character of the approach can lead to broken-symmetry
solutions, showing the electron localization directly in
the electron density~\cite{muller1996,yannouleas1999}. This
method also indicates a clear ``phase transition''
point ($N$-dependent $r_s$) where the crystallization 
occurs. However, going beyond Hartree-Fock to the 
exact diagonalization results makes the situation 
more complicated: There is no clear phase transition
in these finite systems, but the localization gradually
becomes stronger when $r_s$ increases~\cite{reimann2000}.

The most studied system in this context is the two-electron quantum dot, 
the so-called ``quantum dot helium''~\cite{pfannkuche1993}, 
which is in some cases exactly
solvable~\cite{taut1994,elsaid1996,gonzalez1996,dineykhan1997,zhub1998}.
Zhub {\it et al.}~\cite{zhub1998} have shown that at low densities 
(weak confinement, small $\omega_0$) the many-particle
excitation spectrum can be described with the
rotation-vibration spectrum of two localized electrons.
We will return to this method in the case of rotating systems in the
next section.

\begin{figure}[h]
\includegraphics[width=0.95\columnwidth]{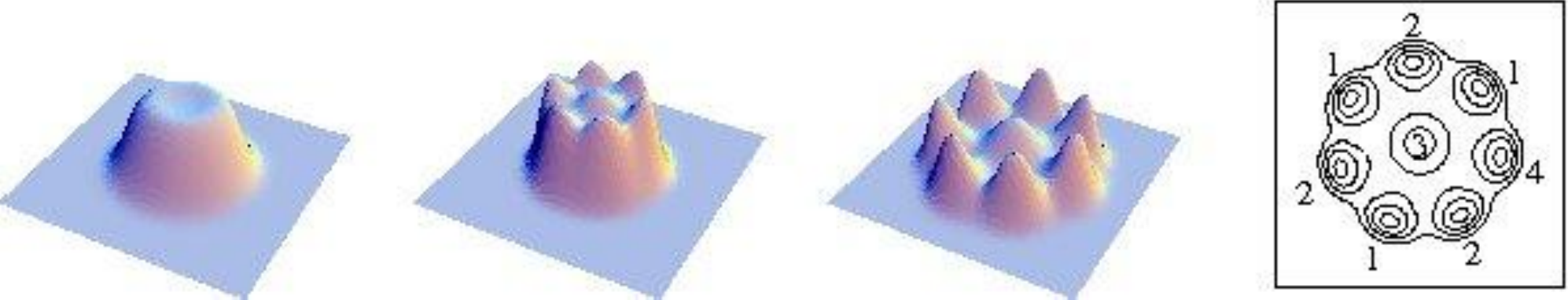}
\caption{Electron density of a four-component quantum dot for 
three different strengths of the confinement frequency, corresponding to
different values of the electron density parameter at the center of
the dot: $r_s=2\ a_0$ (left), $6\ a_o$ (center), and
$14\ a_0$ (right). 
The localization in the 
multicomponent LDA is made possible by the fact 
that the neighboring localized 
electrons belong to different components as 
indicated by the numbers in the contour plot.}
\label{multi}
\end{figure} 

As mentioned earlier, the LDA does not support
electron localization due to the incomplete cancellation
of the direct and exchange Coulomb interaction.
The introduction of spin-dependence (LSDA)
increases slightly the tendency of localization. 
If more internal degrees of freedom are included,
as it could be done in the case of multi-valley semiconductors,
the localization of electrons is expected to happen also
in the local approximation. 
This possibility was studied by 
K\"arkk\"ainen {\it et al.}~\cite{karkkainen2004}.
Figure \ref{multi} shows the localization of 8 electrons 
in a multicomponent electron system when the 
confinement becomes weaker.

\section{Rotating systems in 2D harmonic oscillator}

\label{sec:vortices1}

A semiconductor quantum dot in the presence of a perpendicular 
magnetic field is a finite-size realization of the 
quantum Hall liquid (QHL), which has been an exciting
system of study for both experimentalists and theorists. 
In fact, when Laughlin~\cite{laughlin1983} suggested his
celebrated wave function for the fractional quantum Hall state, he
used exact diagonalization calculations for a finite
quantum dot to test his ingenious Ansatz. 
Since then, it has 
been one of the systems used to mimic also the infinite
systems in many-particle physics of QHL. 

The conductance measurements through a quantum dot show
that as a function of the magnetic field $B$, the conductance
has rather complicated oscillations at small $B$-values~\cite{oosterkamp1999}.
These are caused by successive changes in the quantum states of the electrons,
characterized by changes in the angular momentum and spin quantum numbers.
However, at a certain field range the oscillations disappear 
and it is believed that the electrons form an integer
QHL (with filling factor one). At this state the electron 
system is fully polarized. 

The density functional theory has been extended to treat
2D electrons in the presence of magnetic fields~\cite{vignale1987}. 
In the so-called current-density functional method 
the exchange-correlation energy of the electrons depends,
in addition to the electron density and polarization,
also on the {\it local current density} of the electrons.
Although the method can be disputed in being not 
uniquely defined in all cases and its functionals are
not well established~\cite{capelle2002,dagosta2005}, 
it has been useful in 
understanding the general 'phase diagram'
of the conductance, and has been successful to suggest
new kinds of symmetry-broken ground states, with localized edge 
states~\cite{reimann1999,koskinen1999b} and 
vortices~\cite{saarikoski2004} (see below) as prominent examples. 

\begin{figure}[h]
\includegraphics[angle=-90,width=0.8\columnwidth]{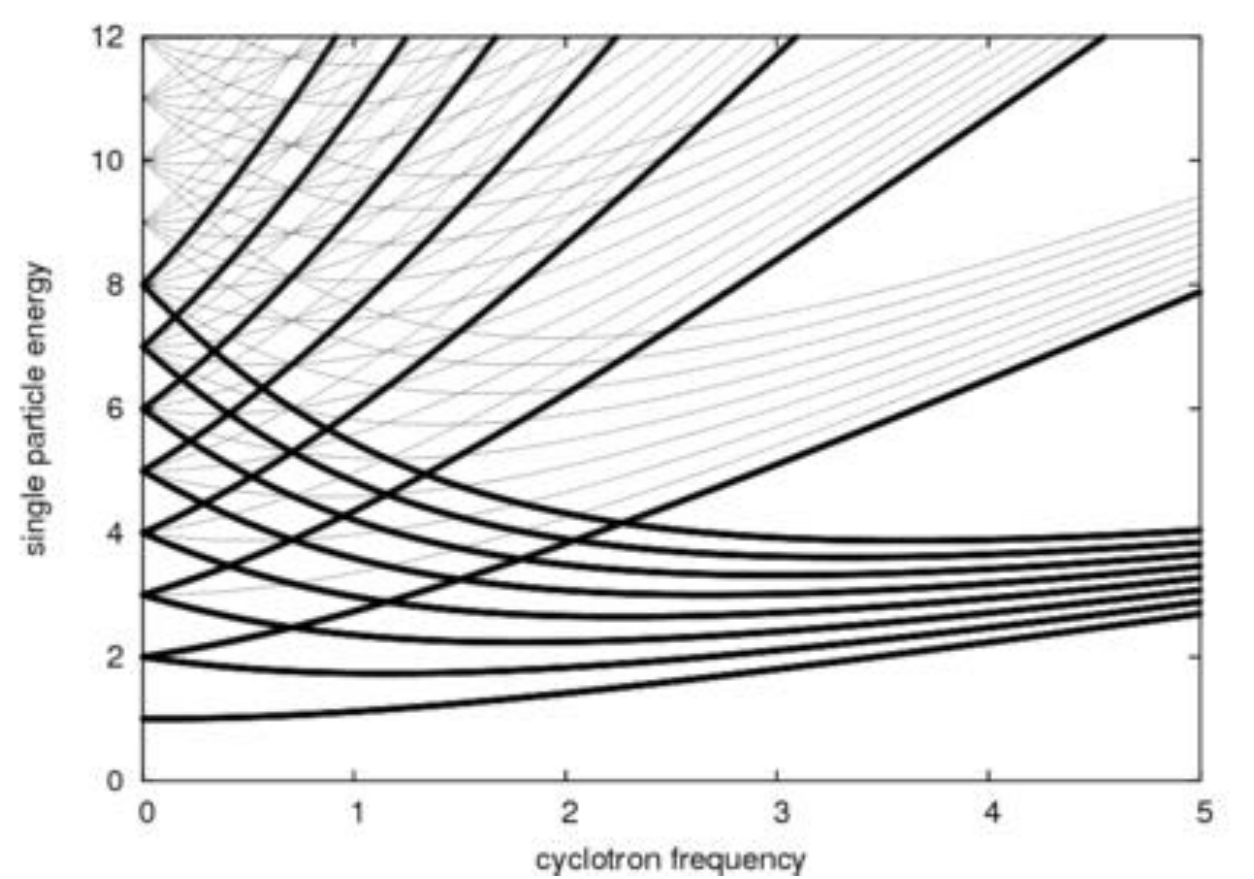}
\caption{Single electron states in 2D harmonic oscillator in a 
perpendicular magnetic field. The levels are plotted as a function
of the cyclotron frequency $\omega_c$. The levels with $n=0$
in Eq.~(\ref{dfstates}) are shown as thick lines.}
\label{darwin}
\end{figure}

The perpendicular magnetic field in the 2D harmonic 
confinement has two effects: It interacts directly
with the magnetic moments of the electrons causing 
a Zeeman term $g\mu_BS_ZB$, and changes the single 
electron kinetic energy from $p^2/2m$ to
$({\bf p}-e{\bf A})^2/2m$. Using the symmetric gauge
for the vector potential, ${\bf A}=(1/2)B(-y,x,0)$, 
and the definition of the cyclotron frequency
$\omega_c=eB/m$ the single-particle Hamiltonian
becomes 
\be
h=\frac{\hbar^2}{2m}\nabla^2 +\frac{1}{2}
m\left(\omega_0^2+\frac{1}{4}\omega_c^2\right)r^2
+\frac{1}{2}\omega_c\hat l,
\label{spH2}
\ee
where $\hat l$ is the $z$-component of the orbital angular
momentum operator. Note that in 2D systems this is the only 
component, and thus, the same as the total angular momentum.
We denote the single-particle angular momentum by $m$ 
and the many-particle angular momentum by $L$.
Clearly, the single-particle problem is exactly
solvable, as discussed already by Fock~\cite{fock1928},
Darwin~\cite{darwin1930}, and Landau~\cite{landau1930}.
The single-particle energies can be written as
\be 
\epsilon_{nm}=\hbar\omega_h(2n+\vert m\vert +1)
+\frac{1}{2}\hbar\omega_cl,
\label{dfstates}
\ee
where $\omega_h=\sqrt{\omega_0^2+\omega_c^2/4}$ and the radial
quantum number is $n=0,\quad 1, \quad ,\cdots$ with
the angular momentum $m$ being an integer.
Figure~\ref{darwin} shows the single-particle states
as a function of the cyclotron frequency (magnetic field).
Only the levels with $\vert l\vert \le 7$ are shown to illustrate
clearly the separation of the levels to different
Landau bands at large values of the magnetic field (large $\omega_c$).
The lowest of these consists of states with $n=0$ and 
$m=0,\quad -1,\quad -2, \cdots$ in increasing order of energy.

Normally, the Land\'e factor $g$ in the Zeeman energy is nonzero.
Consequently, in a strong magnetic field the electron
system will polarize. In this case, the non-interacting
electrons fill the $N$ lowest energy states of the lowest
Landau level (LLL). 
The single-particle states of the LLL are simply
\be
\psi_m(r,\phi)=C_mr^le^{-r^2/2\ell_h^2}e^{il\phi},
\ee
where $C_m$ is the normalization constant and
$\ell_h=\sqrt{\hbar/m\omega_h}$ is the effective oscillator length.
In the theory of QHL it is customary to describe the electron
coordinates by a complex number $z=x+iy$, where
$x=r\cos \phi$ and $y=r\sin \phi$. 
The ground state of polarized noninteracting electrons is 
a Slater determinant formed from the $N$ lowest
single-particle states. Conveniently, it can be written
(in the complex plane) as
\be
\Psi_{MDD}(z_1,z_2,\cdots,z_N)=
\prod_{i<j}^N (z_i-z_j)e^{-\sum \vert z_k\vert^2/2\ell_h^2},
\ee
where the normalization is omitted.
This state is called the maximum density droplet (MDD) and
is the finite-size analog of an infinite integer QHL.
Note that the state is antisymmetric and has the total
angular momentum $L_{MDD}=N(N-1)/2$.

The electron density of the MDD is 
constant inside the dot, as illustrated in Fig.~\ref{edge},
(note that the  density of a single Slater determinant is simply
$n(r)=\sum \vert\psi_m\vert^2$).
In the case of non-interacting, polarized electrons 
the increase of the magnetic field does not change the
structure of the system, but the MDD becomes smaller and
smaller since $\ell_h$ decreases when $B$ (or $\omega_c$)
increases. Before we turn to the much more interesting  
case of interacting electrons, let us note a few facts about 
the excited states of non-interacting fermions in the LLL.

The only way to excite electrons (for a fixed $B$ or $\omega_c$)
in the LLL is to increase the single-particle angular momenta
$m$ such that the total angular momentum increases by $\Delta L$.
This gives an excitation energy of $\Delta E=\hbar\omega_h L$.
However, the degeneracy of the state is in general large,
since there are many ways to distribute 
the single-particle states in the LLL so that the total
angular momentum is $L_{\rm MDD}+\Delta L$.  
The wave function can be written in the complex coordinates
as
\be
\Psi_{\delta L}=P(z_1,z_2,\cdots,z_N)
\prod_{i<j}^N (z_i-z_j)e^{-\sum \vert z_k\vert^2/2\ell_h^2},
\ee
where $P$ is any homogeneous {\it symmetric}
polynomial of order $\Delta L$. The proper antisymmetry is 
provided by the determinant $\prod (z_i-z_j)$.

\subsection{Interacting electrons in the LLL}

\begin{figure}[h]
\includegraphics[width=0.8\columnwidth]{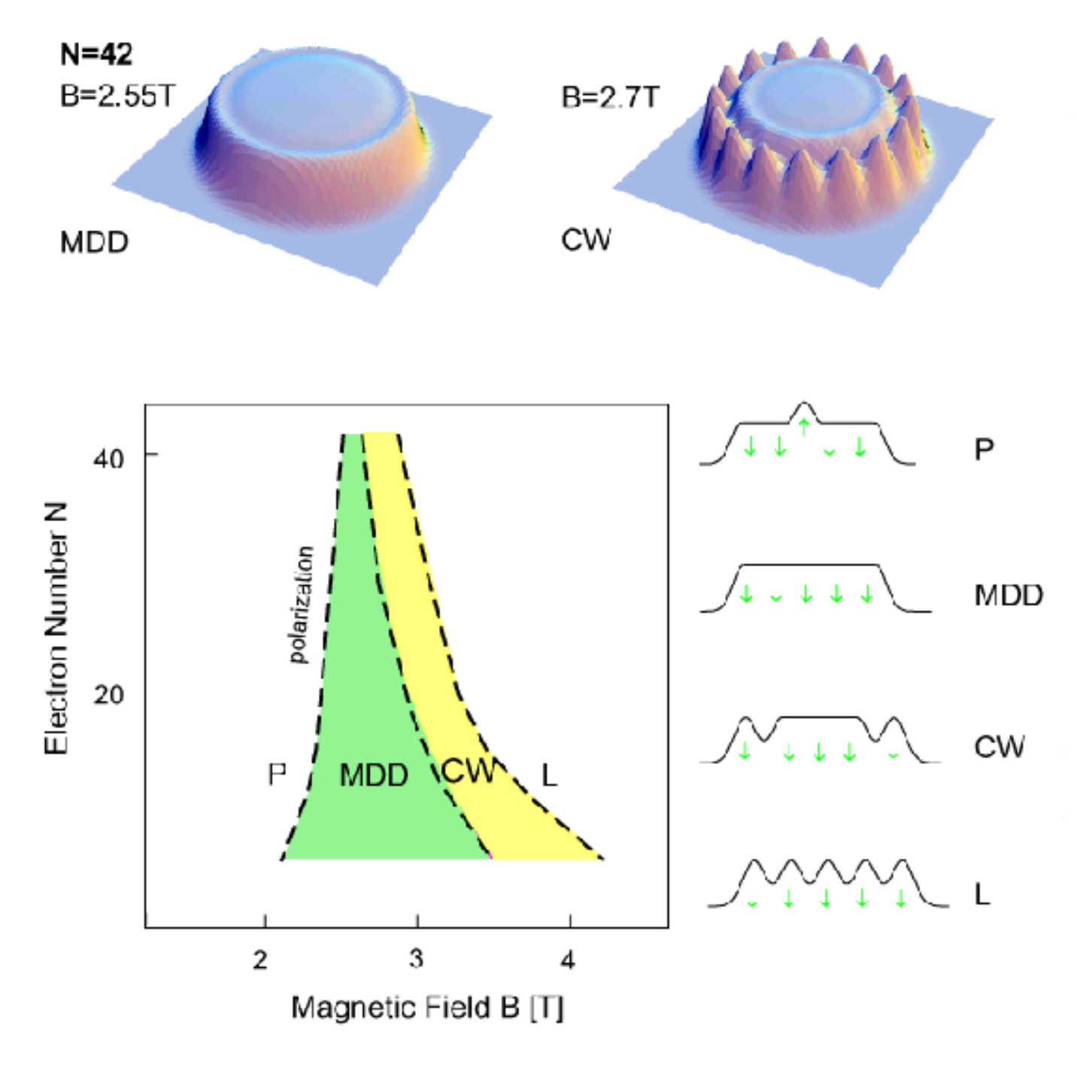}
\caption{``Phase diagram'' for electrons in a harmonic confinement 
in the presence of a magnetic field: P denotes the region of 
where the polarization happens, MDD is the maximum density droplet,
CW is the region of the edge reconstruction, and L denotes the 
high-field region where electron localization sets in. Schematic 
densities and spin configurations of the different regions are shown
at the right. The two figures on top show calculated electron
densities for 42 electrons in the region of the MDD (left) and
CW (right).
The confinement strength was set to $\hbar \omega = 4.192 N^{-1/4} $meV, 
corresponding to typical GaAs values.}
\label{edge}
\end{figure}

The electron-electrons interactions can be included at different 
levels of approximations. The current-density functional theory in the LSDA
takes into account the interactions on a  mean-field level and
allows to include the magnetic field as described above. 
Using the material parameters ($m^*$ and $\epsilon$)
of GaAs, Reimann {\it et al.}~\cite{reimann1999} showed, in agreement
with the experiments, that for each electron number there exists
a region where the ground state is the maximum density droplet. 
This droplet slightly shrinks with increasing magnetic field. 
The ``phase diagram'' shown in  Fig.~\ref{edge} demonstrates how this 
region of MDD ground states becomes narrower when
the number of electrons in the dot increases. (Here, the average electron 
density in the dot was chosen to be approximately constant, 
setting the confinement strength to $\hbar \omega = 4.192 N^{-1/4} $meV, 
which corresponds to a typical value for GaAs).
When the magnetic field becomes too large, the MDD breaks down. 
At large electron numbers, this begins from the 
surface of the droplet. A ring of electrons separates from the 
inner, still compact, droplet.  
The results current-density functional calculations suggest that 
in this split-off ring, the electrons are localized~\cite{reimann1999},
as shown in Fig.~\ref{edge}. 
Again, this broken internal symmetry was 
disputed in the literature.  However, 
calculations based on other many-particle methods
have shown similar localization tendency of this
so-called Chamon-Wen edge in the correlation 
functions~\cite{goldmann1999,toreblad2006}.

\begin{figure}[h]
\includegraphics[width=1.0\columnwidth]{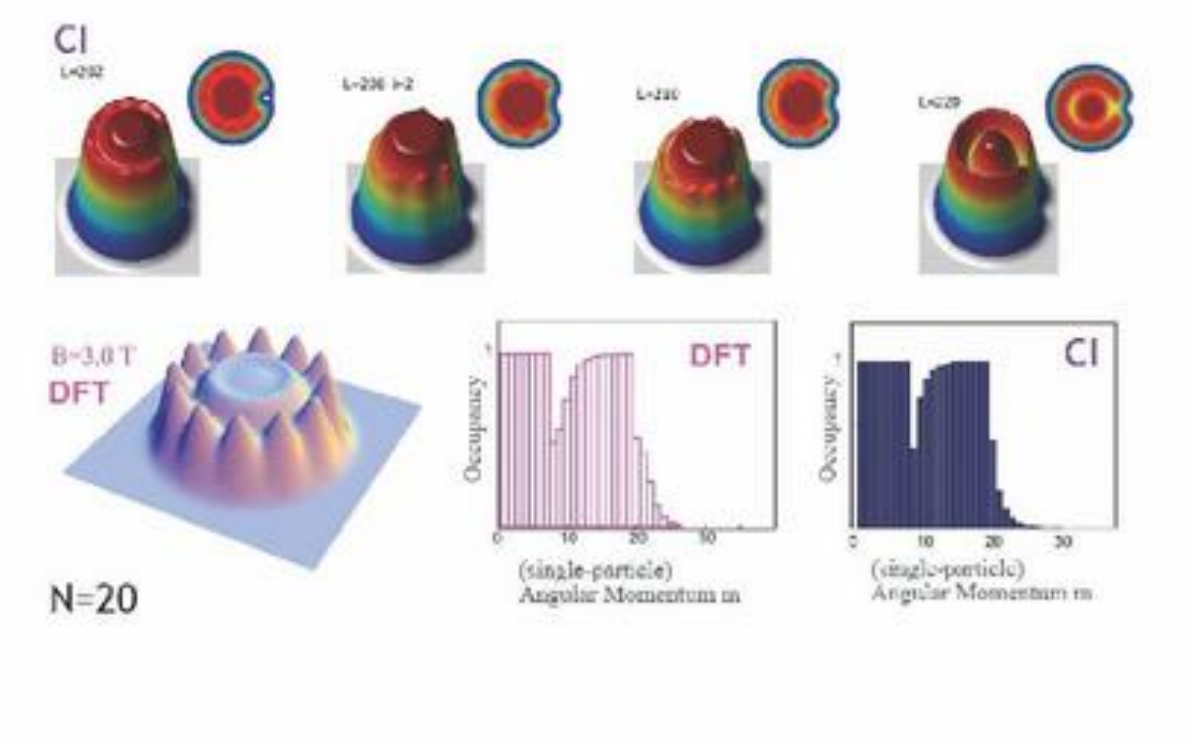}
\caption{Comparison of CI correlation functions (upper panel) and mean-field 
densities, 
for a 20-electron quantum dot at high rotation, or equivalently, 
strong magnetic fields. 
The CI results were obtained for rotation in the Lowest Landau Level (LLL)
only,  for fixed angular momentum as specified. The mean-field result (lower panel,
left) was calculated in CSDFT, at an effective magnetic field of 
$B=3.0T$ ($r_s=2{a_0}^*$).
The two plots at the right-hand side of the lower panel compare the 
occupancies of the single-particle levels in the LLL, characterized by their 
single-particle angular momentum $m$. 
}
\label{reconstruct}
\end{figure} 
Fig.~\ref{reconstruct} compares the correlation functions obtained from 
the CI calculations, to the corresponding result in mean-field 
current spin density functional theory (CSDFT). As an example, we here 
chose the 20-electron quantum dot at high rotation, or equivalently, 
strong magnetic fields. The broken-symmetry along the 
so-called Chamon-Wen edge is reproduced in the CI correlations. 
The occupancies of the single-particle levels in the LLL, characterized
by their single-particle angular momentum $m$, agree remarkably well, 
demonstrating the success of CSDFT in describing the correlated electronic 
structure at strong magnetic fields. 

\subsection{Rotation versus magnetic field}

A magnetic field applied to the 2D harmonic
oscillator leads to the simple Hamiltonian Eq.~(\ref{spH2}).
For a fixed angular momentum $l$ the last term of the
Hamiltonian is a constant and the solutions are
the harmonic oscillator energies and wave functions
for the effective confinement 
$\omega_h=\sqrt{\omega_0+\omega_c/4}$.
This is an important notion:  we can equivalently study the 
rotational spectrum of the harmonic oscillator.
For simplicity, we will now neglect the Zeeman effect,
i.e. the direct interaction between the electron spins and
the magnetic field, $g\mu_BS_zB$. (In fact, in semiconductors
the effective Land\'e factor $g$ can be reduced to zero).

\begin{figure}[h]
\includegraphics[width=0.5\columnwidth]{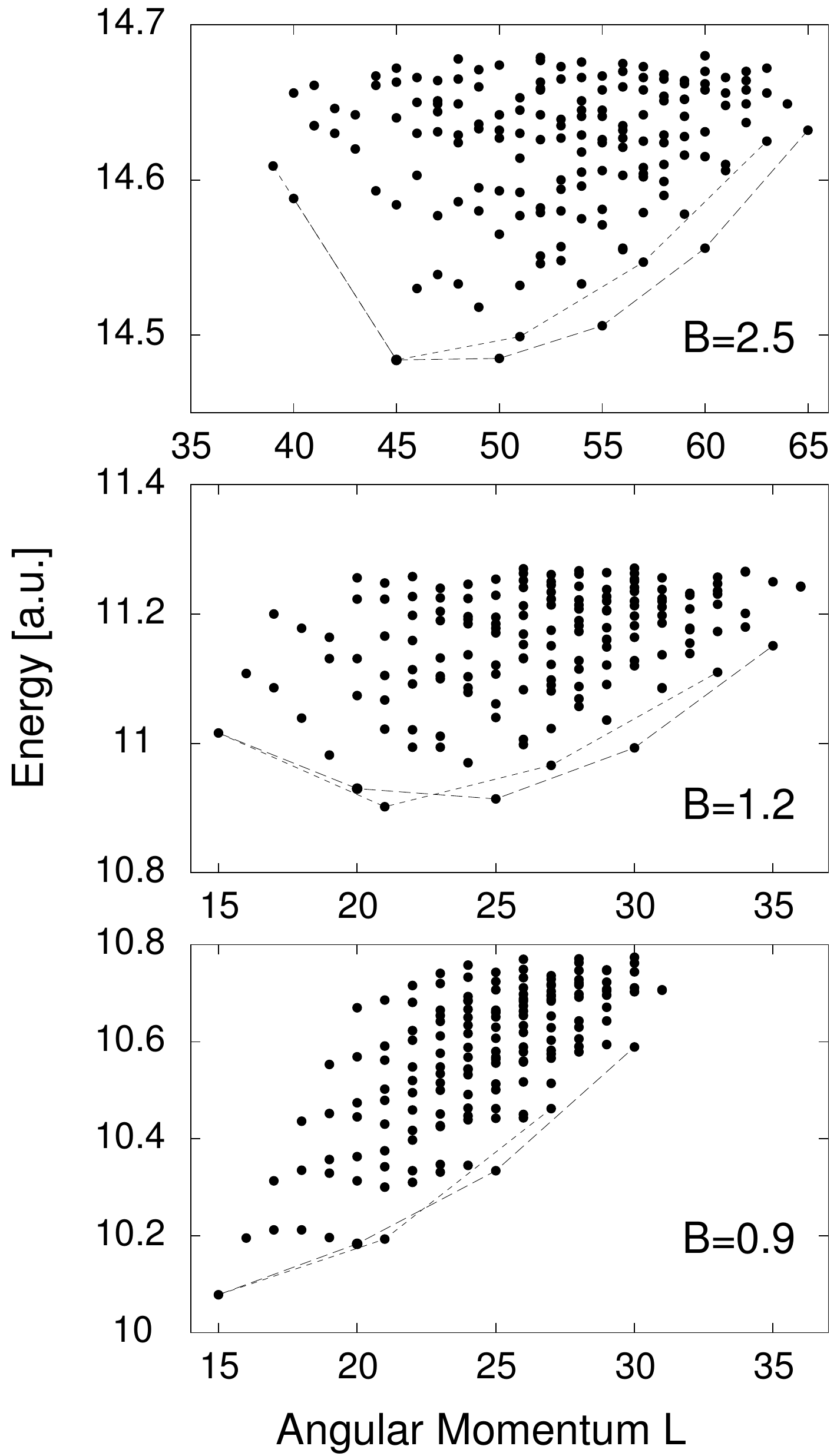}
\caption{Many-particle energy spectrum as a function of the
total angular momentum for three different values of the 
magnetic field (given in atomic units). From Ref.~\cite{manninen2001}.}
\label{bfield}
\end{figure}

Similarly, for the many-particle system,
even when the interactions are included,
the effect of the magnetic field for a fixed $L$
is to increase the strength of the confinement.
Clearly, the Hamiltonian 
can be written as
\be 
H=\sum_i
\left(-\frac{\hbar^2}{2m}\nabla_i^2+\frac{1}{2}m\omega_h^2r_i^2\right)
+\sum_{i\ne j}v(\vert{\bf r}_i-{\bf r}_j\vert) 
+\frac{1}{2}\omega_c\hat L,
\ee
where now $\hat L$ is the total angular momentum.
Again, if the total angular momentum is fixed, the last
term reduces to a constant: in the case of a
2D harmonic confinement, the effect of the magnetic field 
is only to put the system in rotation, and to increase the
strength of the confinement. In Fig.~\ref{bfield} the results
of exact diagonalization for six electrons are shown for three
different strengths of the field. Clearly, the relative structure
of the spectra is very similar, and the effect of the field is only
to tilt the spectrum towards higher angular momenta and to 
determine the energy scale via $\omega_h$.
The rotational spectrum alone
reveals all the effects the magnetic field can have
(apart from the simple Zeeman term), making the direct
comparison to other rotating systems (like for example, 
cold, atomic quantum gases) meaningful.

\subsection{Localization of particles at high angular momenta} 

We will now study the interacting system in a rotational state with
a very high angular momentum. First, let us consider fermions.
For small particle numbers the exact diagonalization technique can be
used with the harmonic oscillator states as the single-particle basis.
When the angular momentum is large all the low energy states are in
the lowest Landau level (LLL) as shown in Fig.~\ref{darwin}, and 
the basis set can thus be restricted to include only the LLL.
With this restriction the matrix size will be finite 
(for a fixed $L$) and for a small particle number no other 
approximations are needed. 

Using the formalism of second quantization, the Hamiltonian
for the polarized electrons (we drop the spin index) is 
\be 
H=\sum_l \hbar\omega_0 l c_l^+ c_l
+\sum_{\{i\}} v_{i_1,i_2,i_3,i_4} c_{i_1}^+
c_{i_2}^+ c_{i_3} c_{i_4}.
\label{h3}
\ee
For a fixed angular momentum $L$ the diagonal term of the 
Hamiltonian gives the energy $\hbar\omega_0 L$ for all 
configurations, thus, just adding a constant. 
The diagonalization of the Hamiltonian is thus reduced to the 
non-diagonal interaction term.
The effect of the confinement frequency $\omega_0$
(or $\omega_h$) is to provide the single-particle basis 
and to determine the energy scale through the interaction
matrix elements $v_{i_1,i_2,i_3,i_4}$.
The many-particle states are completely independent of
the confinement strength, when only the LLL is included
in the basis. When studying the rotational energy spectrum, 
it is thus customary to plot 
the interaction energy, instead of the total
energy. When the angular momentum of the system increases,
the systems expands and the interparticle interactions 
decrease. The interaction energy then decreases with increasing
angular momentum, as seen in the figures below.

\begin{figure}[h]
\includegraphics[angle=-90,width=0.6\columnwidth]{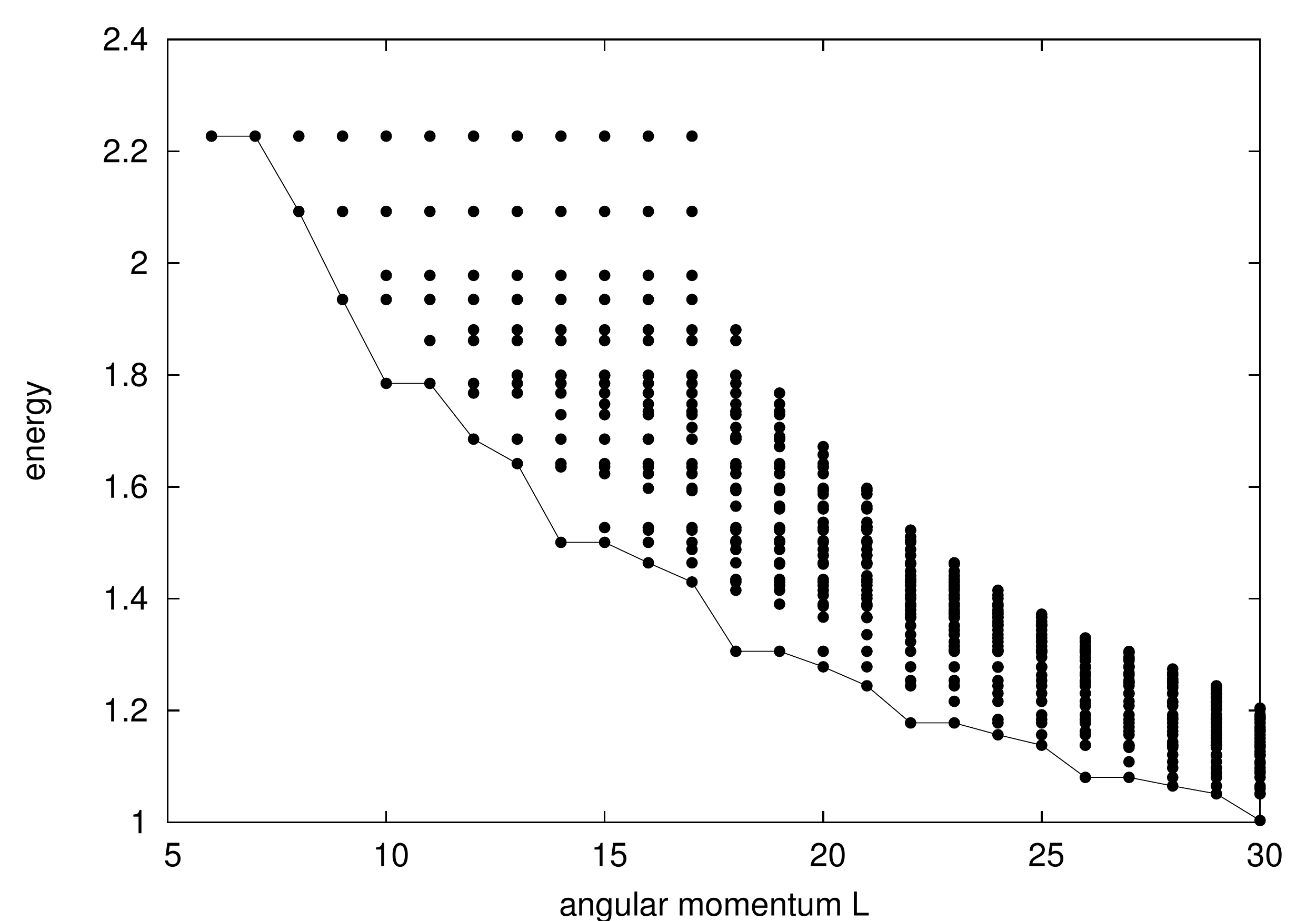}
\caption{Many-particle energy spectrum 
(the interaction energy) for 4 electrons in
a harmonic confinement as a function of the angular momentum.
The lowest energy states are connected with a line to illustrateal
the period of four.}
\label{spect4f}
\end{figure} 

Figure~\ref{spect4f} shows the energy spectrum calculated 
for four electrons as a function of the total angular momentum. 
Two features are distinct. 
First, each appearing new energy is repeated for all higher
angular momentum values. This is due to the center-of-mass
excitations. As discussed in Section~\ref{harmonic}, the 
center-of-mass motion separates from the internal motion, and its
excitation energy is $\hbar\omega_0n$.
In the LLL, each center-of-mass excitation increases the angular
momentum by $\Delta L=1$, but since this does not change the interaction
energy, it remains constant. 

The second important feature of Fig.~\ref{spect4f} is the
periodic oscillation of the lowest energy state as a function
of the angular momentum. These 'yrast' states, i.e. the states 
with highest possible angular momentum at a fixed energy, 
are connected with a continuous line in the Figure. (Actually, 
the name '{\it yrast}' comes from Swedish language for ``the most dizzy'', 
and originates from nuclear physics~\cite{grover1967,BM}. 
The periodic oscillation,
which becomes more distinct when the angular momentum increases,
is a caused by localization of the electrons~\cite{maksym1996,maksym2000}.
Assuming that the electrons are localized in a Wigner molecule,
which in the case of four electrons has the geometry of a square, 
the rigid rotation
of this molecule can be quantized. The symmetry requirements 
of the total wave function allow only every fourth 
angular momentum for a rigid rotation~\cite{maksym2000}. 
These $L$-values correspond precisely to the low-energy 
cusps of the yrast line.
The points in between can not be pure rigid rotations and 
must be other internal excitations. One possibility are, for example, 
the vibrational modes of the Wigner molecule.

To understand the rotation-vibration spectrum
of the Wigner molecules, one can use methods familiar 
from molecular physics. The corresponding energy
levels are 
\be
E_{\rm cl}=E_{\rm cl}^0
+\frac{L^2}{2I}
+\sum_k \hbar\omega_k(n_k+\frac{1}{2})
 +\hbar\omega_0(n_o+1) 
\label{eclass}
\ee
where $I=\sum mr_i$ is the moment of inertia of the molecule,
$\omega_k$ the vibration frequencies, and the last term gives
the energy of the center-of-mass motion. 
The difference between the Wigner molecule and a normal molecule
is that in the former case the Coriolis force is essential for
determining the vibrational frequencies. In practice, they
have to be determined in a rotational
frame~\cite{maksym1996,nikkarila2007}.
Another important difference is the drastic expansion of the 
Wigner molecule as a function of the angular momentum. This
causes not only the decrease of the vibration frequency, but
also the increase of the moment of inertia.

For four electrons, the vibrational modes can be solved
analytically~\cite{nikkarila2007} and the resulting energy spectrum 
can be constructed by considering which combinations of vibrational
modes and rotational states can be used to construct an antisymmetric
state. This can be done with the help of  
group theory~\cite{maksym1996,tinkham1964}.
Figure~\ref{spect4f} shows part of the rotational spectrum,
computed with exact diagonalization of the quantum mechanical
system.  
It is compared to the spectrum obtained from 
Eq.~(\ref{eclass}), i.e. using classical mechanics 
and group theory. The figure shows an excellent agreement between 
the spectra. This demonstrates clearly that at such high angular momenta,
the four-electron system is just a vibrating Wigner
molecule of localized electrons.

\begin{figure}[h]
\includegraphics[width=0.7\columnwidth]{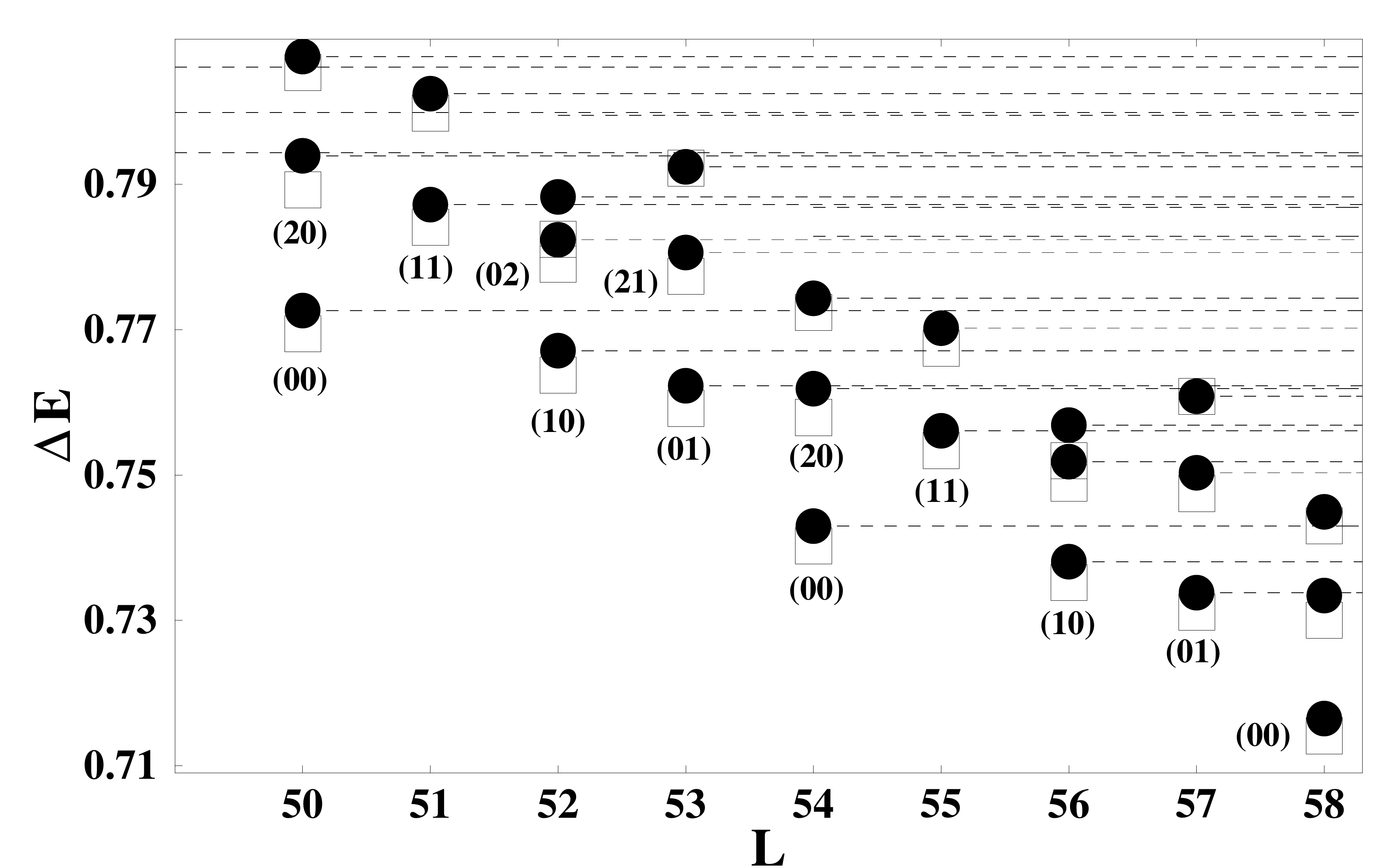}\break
\includegraphics[angle=-90,width=0.7\columnwidth]{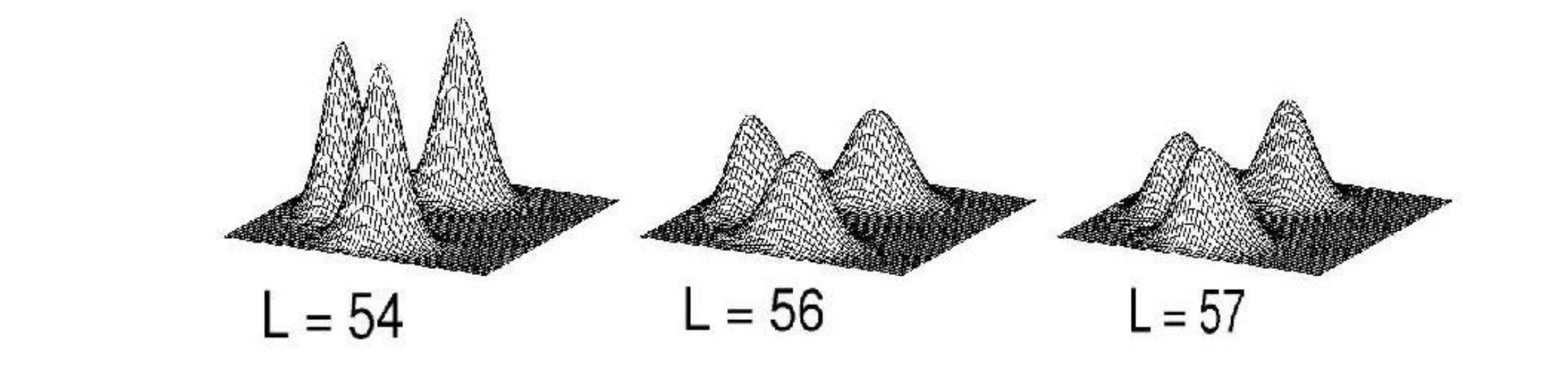}
\caption{Many-particle energy spectrum 
(the interaction energy) for 4 electrons in
a harmonic confinement as a function of the angular momentum.
Solid points: Exact diagonalization; Squares: Model Hamiltonian
(the numbers indicate the vibrational state in question).
The lower panel shows the pair correlation functions for some
yrast states.
}
\label{spec4cl}
\end{figure}

The pair correlation functions shown in the lower panel of Fig.~\ref{spec4cl}
further support this conclusion. For the cusp states 
(as here at $L=50, 54, 58$ for $N=4$), 
the pair correlations show clearly the localization of the 
electrons in a square geometry, while the points in between
these angular momenta  
reflect the properties of the two different vibrational states. 

Finally, let us discuss how this relates to the fractional quantum
Hall effect. Laughlin~\cite{laughlin1983} showed already in his pioneering 
work that the maximum amplitude of the many-electron state in the
fractional QHL
\be
\Psi_{q}(z_1,z_2,\cdots,z_N)=
\prod_{i<j}^N (z_i-z_j)^qe^{-\sum \vert z_k\vert^2/2\ell_h^2},
\label{laughlin}
\ee
is obtained when the electrons are localized
at their classical equilibrium positions. In the wave funtion above, 
$q$ is an odd integer (the filling fraction of the LLL is $\nu=1/q$).
The localization becomes more pronounced 
when $q$ increases~\cite{manninen2001}.
In the region where the true Wigner crystal is formed, the
above wave function is not any more accurate.
In small systems, however, already in the region of 
$\nu=1/3$ ($q=3$) the exact energy spectrum shows the periodic oscillation 
of the yrast spectrum caused by the electron localization
(see Fig.~\ref{spect4f}). The spectrum in Fig.~\ref{spec4cl}
is from the region $\nu\sim 1/9$.

\begin{figure}[h]
\includegraphics[width=0.8\columnwidth]{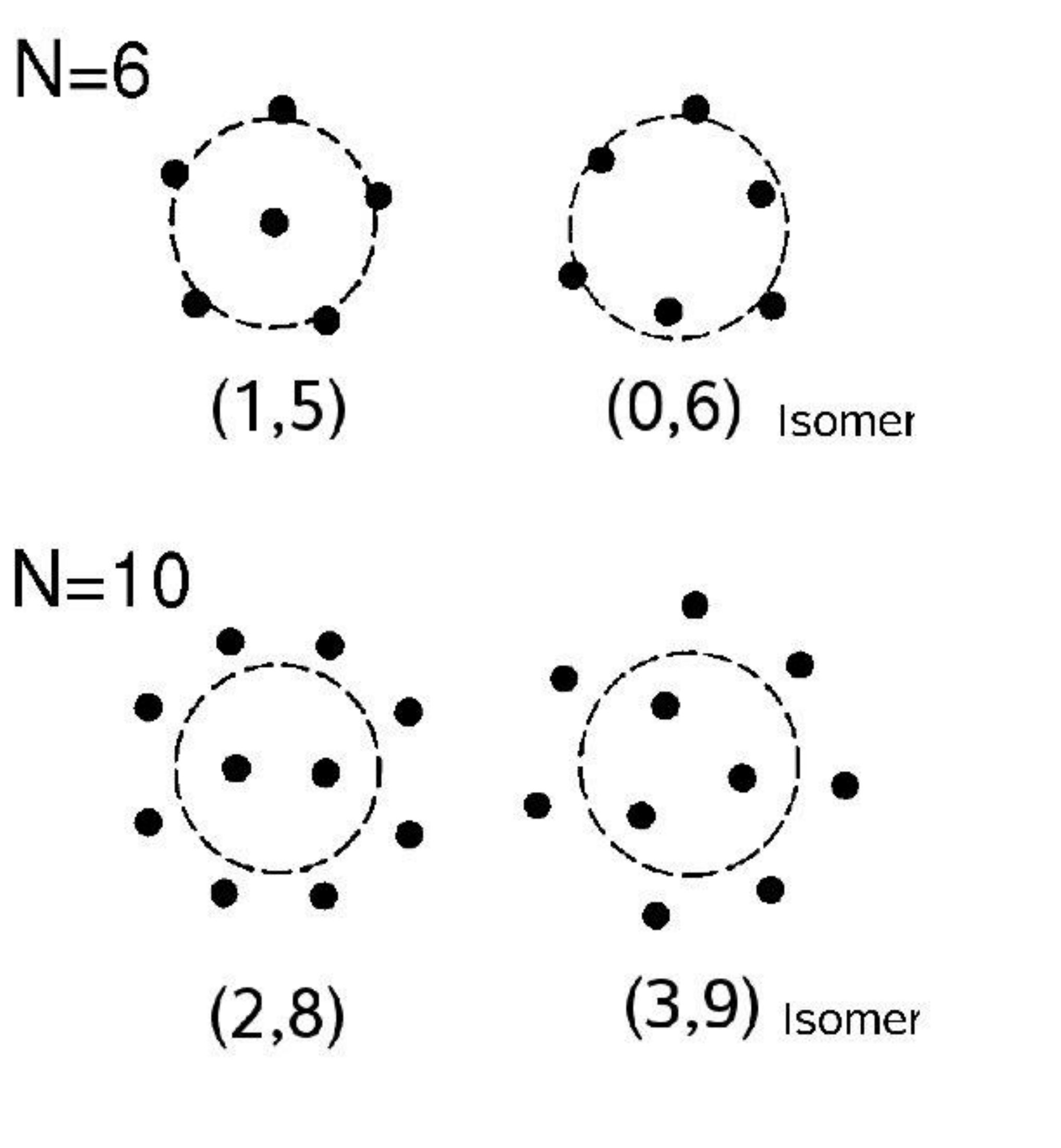}
\caption{Classical electron positions for small particle numbers, 
$N\le 10$, in a parabolic well. 
After Bolton and R\"ossler~\cite{bolton1993}.}
\label{wigclass}
\end{figure}

The classical geometry of the localized electrons depends on their
number~\cite{bedanov1994}. Generally, the electrons tend to form concentric
rings. Up to five electrons they form a single ring, but for
six electrons the ground state is a five-fold ring with one
electron at the center, as schematically shown in Fig.~\ref{wigclass}. 
This figure displays the  classical equilibrium positions, for the example 
of six (upper panel) and ten electrons (lower panel), respectively 
(after Bolton and R\"ossler~\cite{bolton1993}.
See also the discussion in Ref.~\cite{reimann2002}).
\begin{figure}[h]
\includegraphics[width=1.0\columnwidth]{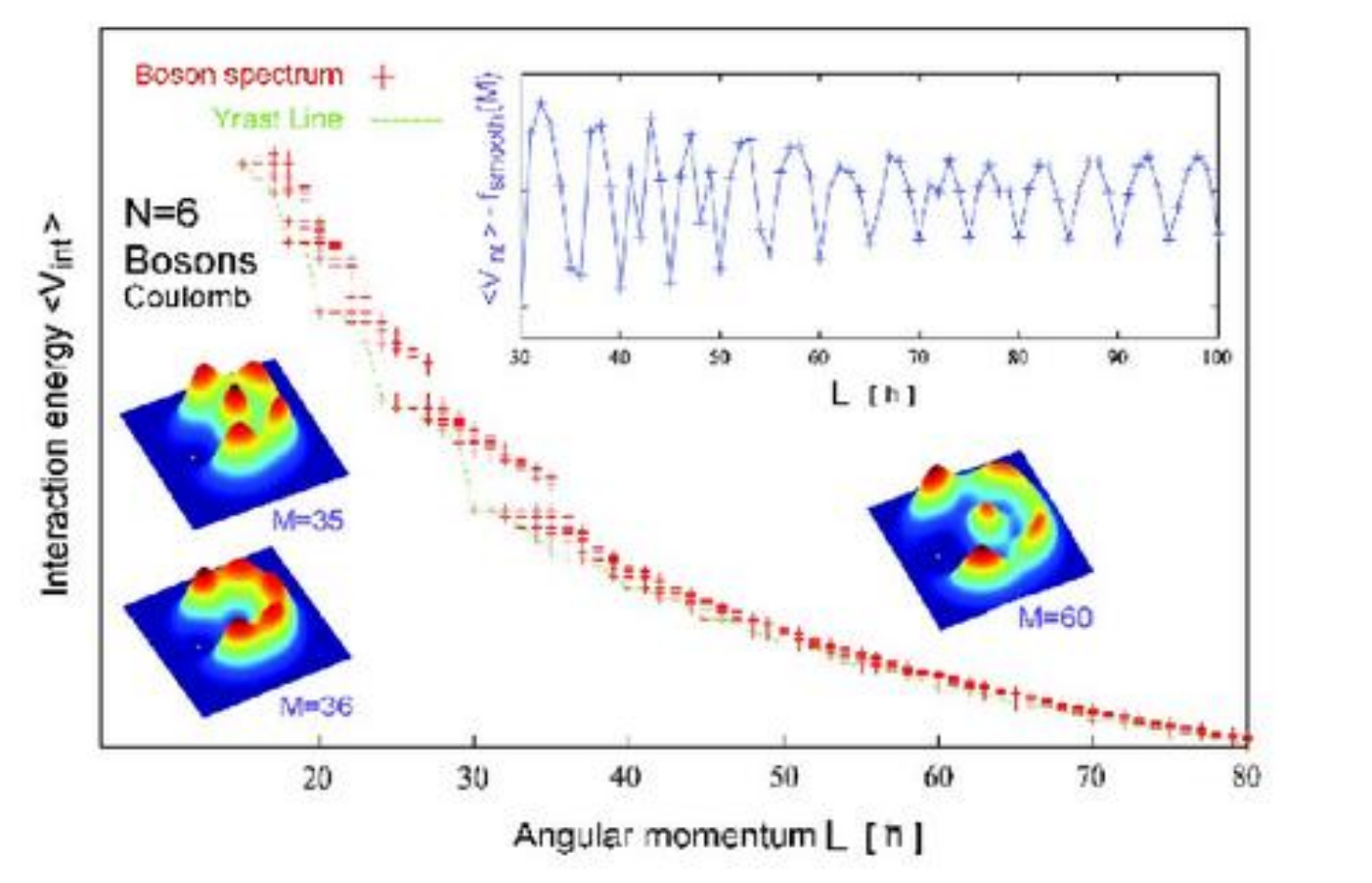}
\caption{Interaction energy of $N=6$ bosons 
as a function of the angular momentum. 
The inset shows the yrast line with a smooth function of angular 
momentum (3rd order polynomial) subtracted from the 
energies, in order to make the oscillations more visible. 
The large-$L$ limit is dominated by a regular oscillation with $\Delta L=5$. 
The pair correlation functions to the left 
clearly demonstrate localization in Wigner molecule geometries
at high angular momenta. 
While at smaller $L$-values, the 
$(1,5)$ and the $(0,6)$ configurations compete, at extreme angular momenta 
fivefold symmetry dominates.}
\label{spect6}
\end{figure}
The localization caused by the highly rotational state is not limited
to electrons in a harmonic confinement, but is a more general phenomenon to
occur for all particles
with long-range interactions. The reason behind is simple: At 
large angular momenta, the system can be described by the classical
rotations and vibrations of Eq.~(\ref{eclass}). The different symmetry
requirements can, however, select different allowed vibration
modes for fermions and bosons (at a given angular momentum).

Fig.~\ref{spect6} shows as an example the interaction energy as a function
of the angular momentum, for six bosons interacting by Coulomb repulsion. 
Subtracting a smooth function of angular 
momentum (3rd order polynomial) from the yrast line, pronounced and regular 
oscillations of period $\Delta L=5$ are visible in the 
large-$L$ limit, originating from the localization into a five-fold ring 
with one boson at the center, just as for the fermion case discussed above. 
This localization is confirmed by the pair correlations, shown as an inset in
the same figure. 
Reimann {\it et al.}~\cite{reimann2006} 
have furthermore shown,  that the energy spectra of 
small numbers of bosons and fermions are nearly identical
at high angular momenta.

The Laughlin wave function, Eq.~(\ref{laughlin}),
is also applicable for bosons. In this case, naturally the exponent $q$ needs
to be even. This suggests that there is a relation between the
boson and fermion wave functions. Since the boson wave function is
symmetric, a proper fermion wave function can be constructed by
multiplying the boson wave function 
with the determinant $\prod (z_i-z_j)$.
Indeed, for the wave functions at large angular momenta this
construction gives excellent approximations for the fermion
wave functions. The overlap between this construction
and the exact fermion wave function for four electrons at
high angular momenta is typically 99 \%~
\cite{borgh_tbp}. 
Note, that this is  
not only true for the rigidly rotating states, but also for states 
with internal vibrations.

Why does the rotational motion localize the particles in 
a harmonic confinement? In the case of electrons with long-range 
Coulomb interactions one could think that this is caused by
Wigner crystallization. When the angular momentum 
increases, the electron cloud expands due to the centrifugal force
and eventually, Wigner crystallization sets in.
However, we have already seen that the external magnetic
field increases the strength of the confinement. In fact, 
the average electron density remains essentially constant when 
localization occurs.

\subsection{Vortices in polarized fermion systems}

\label{sec:vortices2}

Vortex formation in type-II superconductors is a well-known
phenomenon~\cite{tinkham1975}. 
When the magnetic field increases, at the first critical
field strength at a given temperature, vortices penetrate the
superconductor forming a regular triangular lattice.
Similarly, in rotating $^3$He vortex formation has been observed
by optical measurements~\cite{manninen1992}. 
Vortex formation in rotating systems has
been considered as a definite signature of superfluidity. 

In the case of semiconductor quantum dots, vortex formation was
discussed theoretically by 
Saarikoski {\it et al.}~\cite{saarikoski2004} using the
current-density functional formalism. Later this was 
confirmed by exact diagonalization
calculations~\cite{toreblad2004,manninen2005}.
The vortices appeared when the magnetic field was increased beyond 
formation of the maximum density droplet, but at field strengths 
below those where the fractional QHL occurs.

\begin{figure}[h]
\includegraphics[width=0.9\columnwidth]{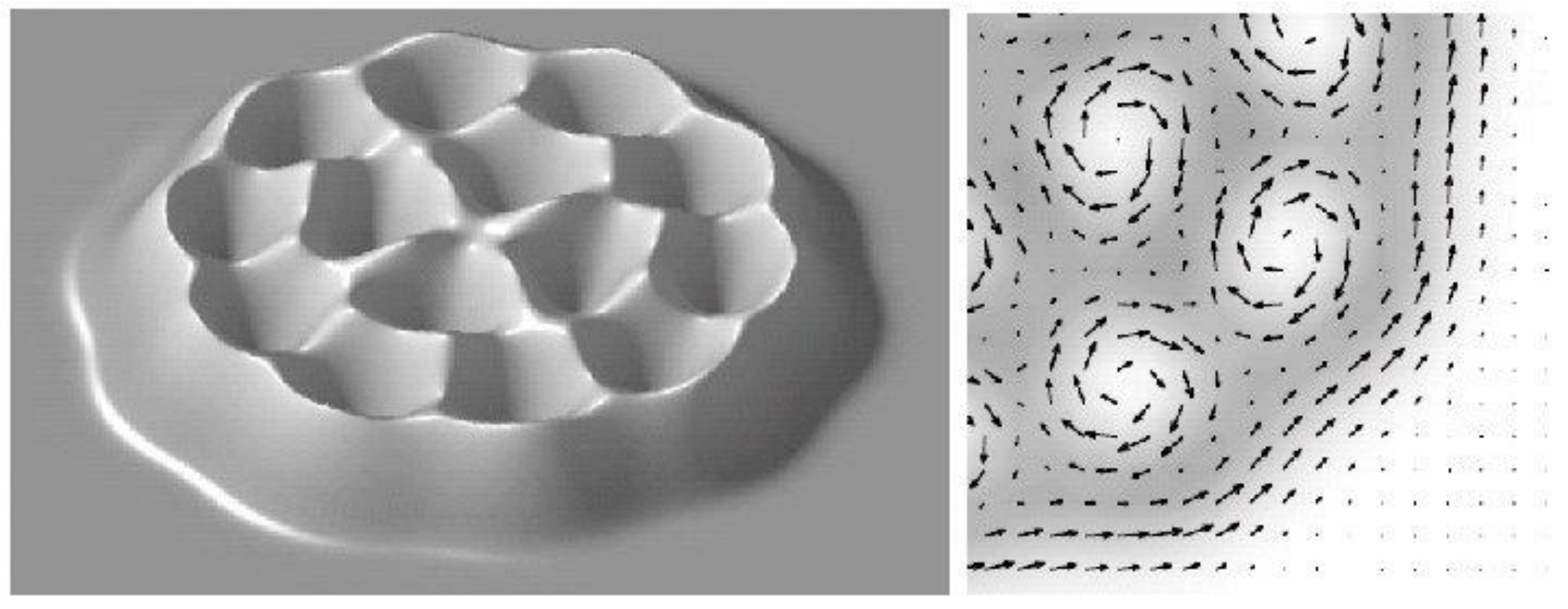}
\caption{Electron density of a 24 electron quantum dot
showing 14 vortices (left) and the corresponding currents (right).
Results from a current-spin-density functional calculation by
Saarikoski {\it et al}. From Ref.~\cite{saarikoski2004}.}
\label{saarikoski}
\end{figure}

Just as for localization of the electrons, as discussed above, 
in mean-field theory the vortices are visible 
directly as  distinct minima in the total electron density, with  the
electron current showing  circulation of the vortex core, as illustrated
in Fig.~\ref{saarikoski}. The vortices seem to localize a in a regular
'molecule', with geometries resembling those observed for the 
finite-size Wigner crystallites discussed above.

\begin{figure}[h]
\includegraphics[width=0.7\columnwidth]{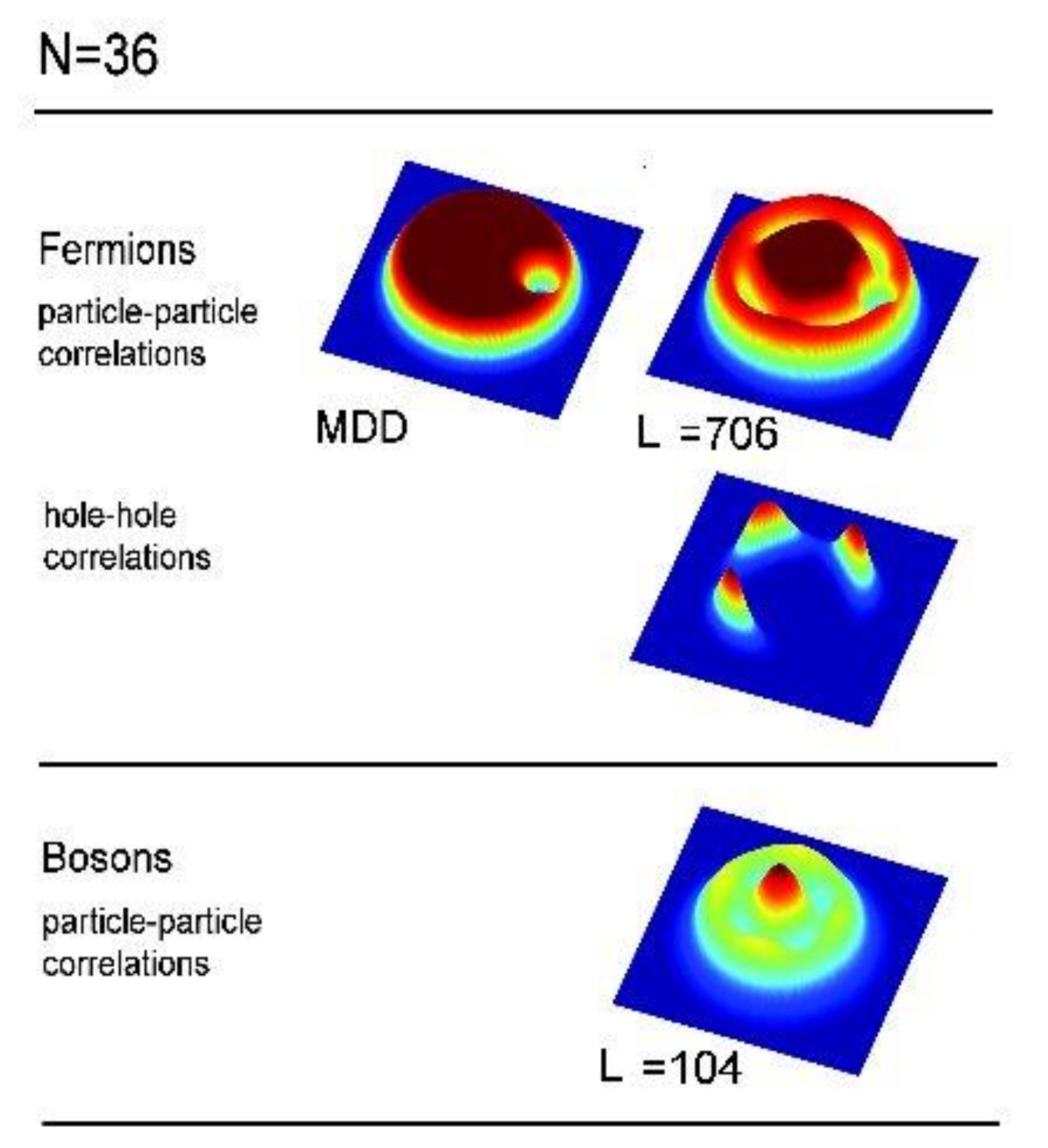}
\caption{Pair correlation functions calculated for 36 electrons.
The {\it upper panel} shows the  electron-electron correlations for the 
MMD, $L=630$ ({\it left}), for particles at 
$L=706$ showing four vortices ({\it right}), and for holes at the same angular
momentum ({\it lower right}). 
The {\it lower panel} shows the corresponding  
correlation function for a  bosonic four-vortex state 
at angular momentum $L=104$. (Note the absence of the exchange hole in the
bosonic case.) From Ref.~\cite{reimann2006b}.}
\label{v4n36}
\end{figure}

To analyze the vortex solutions gained by the exact diagonalization 
method, is not an easy task~\cite{reimann2006b}. 
Naturally, for the exact solution of the many-body Hamiltonian, the total
density is circularly symmetric and one has to study correlation
functions -- just as explained above for the case of Wigner localization. 
Figure~\ref{v4n36} shows the electron-electron
pair correlation for 36 electrons at a highly rotational state. Clearly, in
addition to the exchange-correlation hole around the reference
electron, there are four distinct minima in the pair correlation. These
are four localized vortices -- the reference electron pins
their position, making them visible. 

Other ways to observe the internal symmetry breaking 
in the exact diagonalization
study are to break the circular symmetry, for example by an 
ellipsoidal confinement~\cite{manninen2001b,saarikoski2005},
or by using perturbation theory~\cite{toreblad2004}. The electron density at
the vortex core is zero, and the phase of the wave function changes
by $2\pi$ when a coordinate is rotated around the vortex core. In the
case of the many-particle wave functions, these characteristics are
difficult to use. 
It was suggested by Saarikoski {\it et al.}~\cite{saarikoski2004} 
to determine the phase change of the many-particle wave function by 
by fixing the positions of $N-1$
coordinates when the $N$'th coordinate is rotated around the vortices
(fixing the other coordinates fixes also the positions of the
vortices). The phase maps created 
in this way~\cite{saarikoski2004,toreblad2004} show that in
addition to the 'free' vortices there is one vortex attached
to each electron. In the language of the QHL, each
electron carries a flux quantum (in the case of fractional QHL with
filling factor $\nu=1/3$, each electron carries three flux quanta).
Electrons with attached flux quanta (or vortices) are also called
composite fermions~\cite{jain1998}.

In the polarized case, there is a
simple way to understand the occurrence of free vortices. 
They are {\it holes}  in the otherwise filled Fermi see, i.e. holes in  
the MDD, where all states up to the single-particle angular momentum
$L_{\rm MDD}=N(N-1)/2$ are filled. When the angular momentum is increased
we create holes (missing electrons) corresponding to small angular
momenta relative to $L_{\rm MDD}$.
Formally, we can can define the creation (annihilation)
operator of a hole as $d^+=c$ ($d=c^+$) and write
the Hamiltonian Eq.~(\ref{h3}) in terms of these, 
\be
\label{sqHh}
H=\sum_i m_i\hbar\omega_0(1-d_i^+d_i)
+2\sum_{i,j,k}\left(V_{ijkj}-V_{ijjk}\right)d_k^+d_i
+\sum_{i,j,k,l}V_{ijkl}d_l^+d_k^+d_jd_i+{\rm constant}.
\ee
Note, that the interactions between the holes are the same as those between 
the particles, but the second term means that the particles do not any
longer move in a strictly harmonic confinement.
Naturally, the solution of this Hamiltonian leads to an 
equivalent result as that of the original Hamiltonian,  
requiring the same computational effort.

We will now show that -- even within a limited single-particle 
space -- the holes localize to a Wigner molecule.
Let us consider $n$ holes in a system with $N$ electrons and restrict
the single-particle basis to its minimum possible value in the LLL, i.e. the
maximum single-particle angular momentum being $l_{\rm max}=N+n-1$.
When the angular momentum of the electron system is
$L_e=L_{\rm MDD}+\Delta L$, the angular momentum of the system of holes is
$L_h=(N+n)(N+n-1)-L_e=2nN+n(n-1)-\Delta N$. For example, for the 
$N=4$ particle system (cf. Fig.~\ref{v4n36}), 
$L_h=66$ corresponds to such high angular momentum
that the quasi-particles (now holes) are
well localized.

\bigskip

\begin{figure}[h]
\includegraphics[width=0.7\columnwidth]{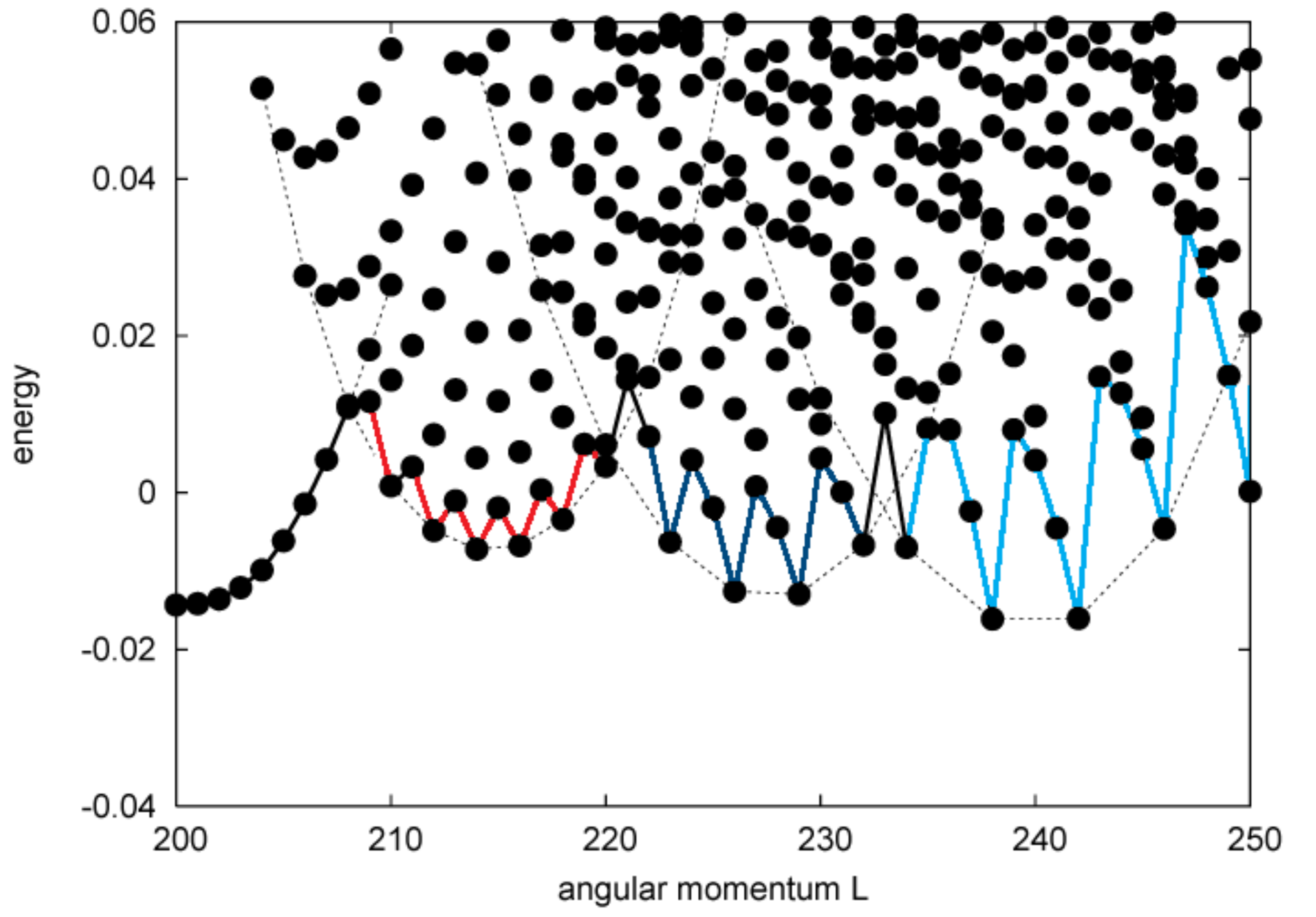}
\caption{Energy spectrum as function of the total angular momentum
for 20 electrons. A smooth function is subtracted from the total
energy to show the oscillations of the yrast line (thick line).
The thin lines show the lowest energy states with 1, 2, 3, and 4
vortices.}
\label{spect20f}
\end{figure}

This localization suggests that
the excitation spectrum can be determined from the ``classical'' rotations and
vibrations, resulting in similar periodic oscillations as found above
for localized electrons at high angular momenta. This indeed is the
case, as shown by 
Manninen {\it et al.}~\cite{manninen2005,reimann2006b}. 
Figure~\ref{spect20f} shows the energy spectrum for 20 electrons as a
function of angular momentum. The yrast-line shows oscillations, in the 
beginning with period 2, followed by oscillations of period 3 and then
period 4 (in units of angular momentum). These regions correspond to
the formation of two, three, and four vortices, respectively.
This means that the main features of the many-particle spectrum at
these angular momenta are determined by the rotation-vibration
spectrum of localized vortices. In Section~\ref{bosons} we will see
that similar oscillations reveal the existence of vortices in rotating
boson systems.

Like for fermions, also in the bosonic case, 
the structure of the bosonic wave function can be understood in
terms of Laughlin-type wave
functions~\cite{toreblad2004,reimann2006b}.
The simplest Ansatz for the single vortex at the center
is the Bertsch-Papenbrock~\cite{papenbrock1999} wave function
\be
\Psi_{\rm 1v}=\prod_i (z_i-z_0)\Psi_{\rm MDD},
\ee
where $z_0=\sum z_i/N$ is the center-of-mass coordinate.
This wave function is a good approximation for the wave function
calculated using only the LLL. For example, for 10 electrons the
overlap between these two states is 
$\vert\langle\Psi_{\rm 1v}\vert\Psi_{\rm exact}\rangle\vert^2=0.90$.
In a large quantum dot, the center of mass can be approximated as
fixed at the origin. Similarly, having $n$ vortices in a ring, we can 
approximate the wave function as 
\begin{eqnarray}
\Psi_{kV} &=&\prod_{j_1}^N (z_{j_1}-ae^{i\alpha_1})\times \cdots\times
\prod_{j_k}^N (z_{j_k}-ae^{i\alpha_k}) \Psi_{MDD} \nonumber\\
&=& \prod_j^N (z_j^k-a^n)\Psi_{MDD}~,
\label{vgs}
\end{eqnarray} 
where $k$ is the number of vortices, $a$ is the distance of the
vortices from the origin and $\alpha_j=2\pi j/k$. 
Clearly, the above wave function does not have a good angular momentum.
Projecting to good angular momentum means collecting out states
with a given power of $a$. We obtain a state
\be
\Psi_{kV}=a^{k(N-K)}{\cal{S}}\left(\prod_j^K z_j^k\right)\Psi_{MDD}
\label{vgen}
\ee
which now corresponds to a good angular momentum $M=M_{MDD}+kK$
(here, $\cal{S}$ symmetrizes the polynomial). The above wave function
corresponds to the most important configuration of the exact wave
function: The $n$ holes are next to each other in consecutive 
angular momenta. Toreblad {\it et al.}~\cite{toreblad2004} called
this state a ``vortex-generating configuration''. (However, the
wave function (\ref{vgen}) does not localize the vortices but 
rather keeps them  de-localized at a distance $a$ from the origin).

\subsection{Vortices in rotating Bose systems}
\label{bosons}

The observation of Bose-Einstein condensation in atomic 
traps once again increased the interest in the many-particle physics
of the  harmonic potential (for a review, see~\cite{pethick2002}). 
The experimental observation of vortex
lattices in rotating systems was a further milestone. By external fields, 
the trap can be made to be three-dimensional or quasi-two-dimensional.
In a highly rotational state, 
the cloud of atoms forms a (quasi two-dimensional) disc, 
with the effective confining potential in the rotating frame being
\be 
V_{\rm ext}(r,z)=\frac{1}{2}
m\omega_{\rm eff}^2r^2
+\frac{1}{2}m\omega_0^2 z
=\frac{1}{2}
m\left(\omega_0^2-\omega_r^2\right)r^2
+\frac{1}{2}m\omega_0^2 z
\ee
where $\omega_r$ is the angular velocity of the
rotation. For large enough $\omega_r$, only the lowest energy state
along the  $z$-direction is occupied.
(Note, that the rotation velocity can not exceed
$\omega_0$). We can then approximate the rotating Bose system as
particles confined in a 2D harmonic trap, and directly compare
to fermions confined in a quantum dot.
\begin{figure}[h]
\includegraphics[width=0.5\columnwidth]{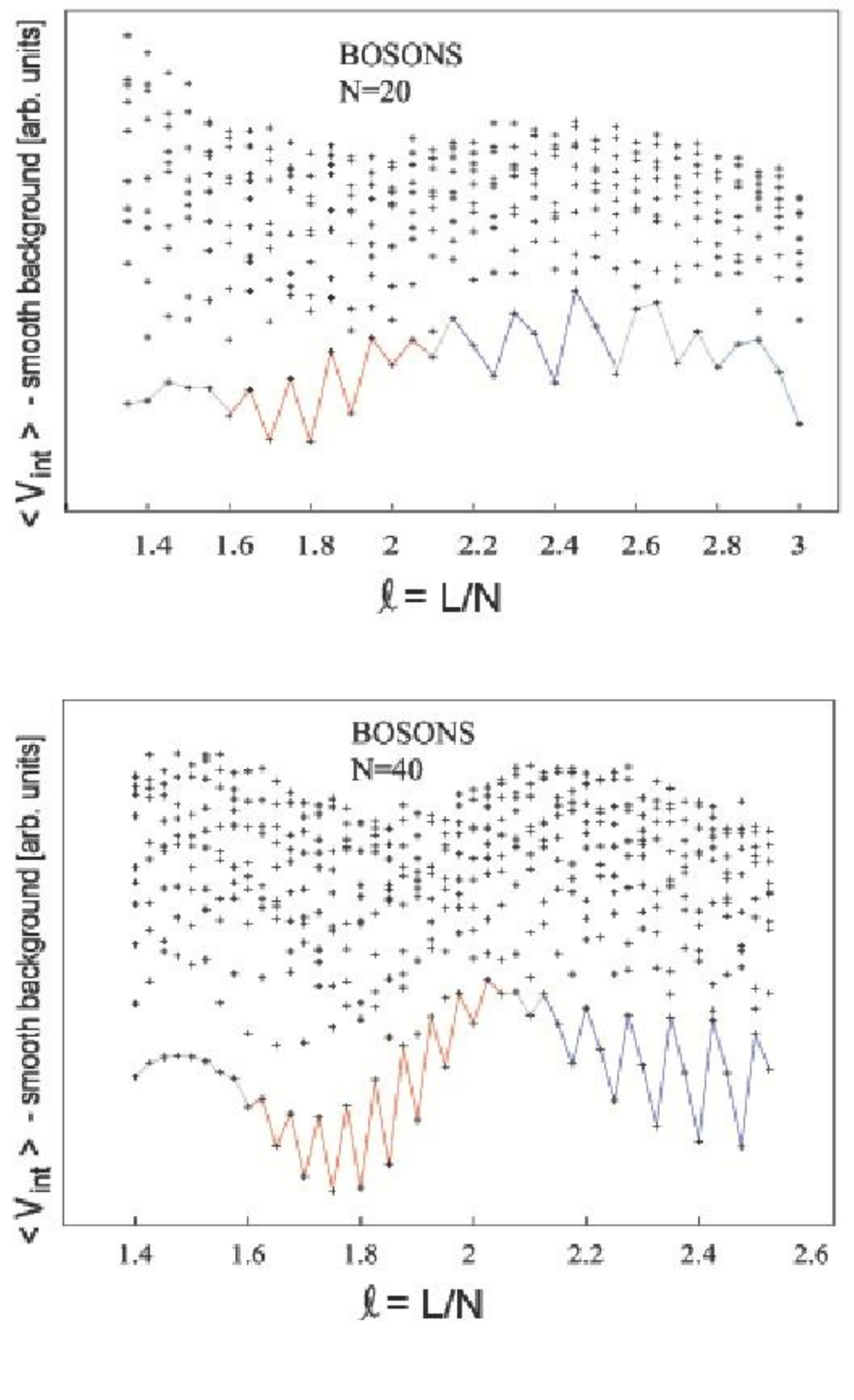}
\caption{Energy spectrum as function of the total angular momentum
for $N=20$ bosons (upper panel) and $N=40$ bosons (lower panel), 
with Coulomb interactions
($L$ is the angular momentum, $N$ the number of bosons).
A smooth function is subtracted from the total
energy to show the oscillations of the yrast line (thick line).}
\label{bv1}
\end{figure} 

The interaction between the atoms in the dilute condensate consists of
individual scattering events which are described by the
scattering length. The {\it contact} interaction, being a standard model 
interaction for cold atom gases, is  written as 
$v({\bf r}_i-{\bf r}_j)=g\delta({\bf r}_i-{\bf r}_j)$, where
$g=4\pi a_s\hbar^2/m$, $a_s$ being the scattering length
(for $s$-wave scattering). In the dilute gas, the total energy
per particle is proportional to the density. Consequently,
in the local density approximation the effective potential
will also be proportional to the density. In a Bose system at
zero temperature, all particles are in the same quantum state,
and the density is simply $\rho({\bf r}=\vert\psi({\bf r})\vert^2$,
where the single-particle wave function $\psi$ is the solution
of the so-called Gross-Pitaevskii equation
\be
-\frac{\hbar^2}{m} \nabla^2\psi({\bf r})
+\frac{1}{2}m\omega_0^2r^2
+g\vert\psi({\bf r})\vert^2
\psi({\bf r})=\epsilon\psi({\bf r}),
\ee 
which is a mean-field equation in close correspondence to the Kohn-Sham
LDA equations for the electron system~\cite{capelle}. The nonlinearity of the
equation makes symmetry-breaking possible. Indeed, for a rotating Bose gas, 
the equation has solutions showing vortex patterns very similar to the 
ones discussed above for the fermion case~\cite{butts1999,kavoulakis2000}.
Figure~\ref{bv1} shows the boson spectra,  with oscillations in the yrast
line resembling to  vortex structures as discussed above in the fermion case.  

For small numbers of bosons in a harmonic potential,
the problem can be solved exactly. 
In the case of weak interactions between the bosons, the basis set can be 
restricted  to the lowest Landau
level. The only difference to the fermion system discussed above is, 
that now the 
wave function has to be symmetric.
For contact interactions, it has been shown that the Bertsch-Papenbrock
Ansatz 
\be
\Psi_{\rm BP}=\prod_i (z_i-z_0)e^{-\sum \vert z_k\vert^2/2\ell_0^2}
=\prod_i (z_i-z_0)\Psi_{\rm BEC}
\ee
is {\it exact} for the state with a single vortex at the
center ($\ell_0$ is now the oscillator length of the 
pure confinement).
This state has total angular momentum $L=N$. Increasing the angular
momentum creates more vortices.  A second vortex appears 
at $L=1.7N$, the third at $L=2.1N$ and the 
fourth at $L=2.8N$~\cite{kavoulakis2000}. 
An approximation for the $n$ vortices in a ring is again the
vortex generating state, Eq.~(\ref{vgs}), where now the 
fermion MDD is replaced by the ground state of the BEC.
But again, as for fermions, 
the exact solution is more complicated and supports
vortex localization in a much more effective way. 

\begin{figure}[h]
\includegraphics[angle=-90,width=0.8\columnwidth]{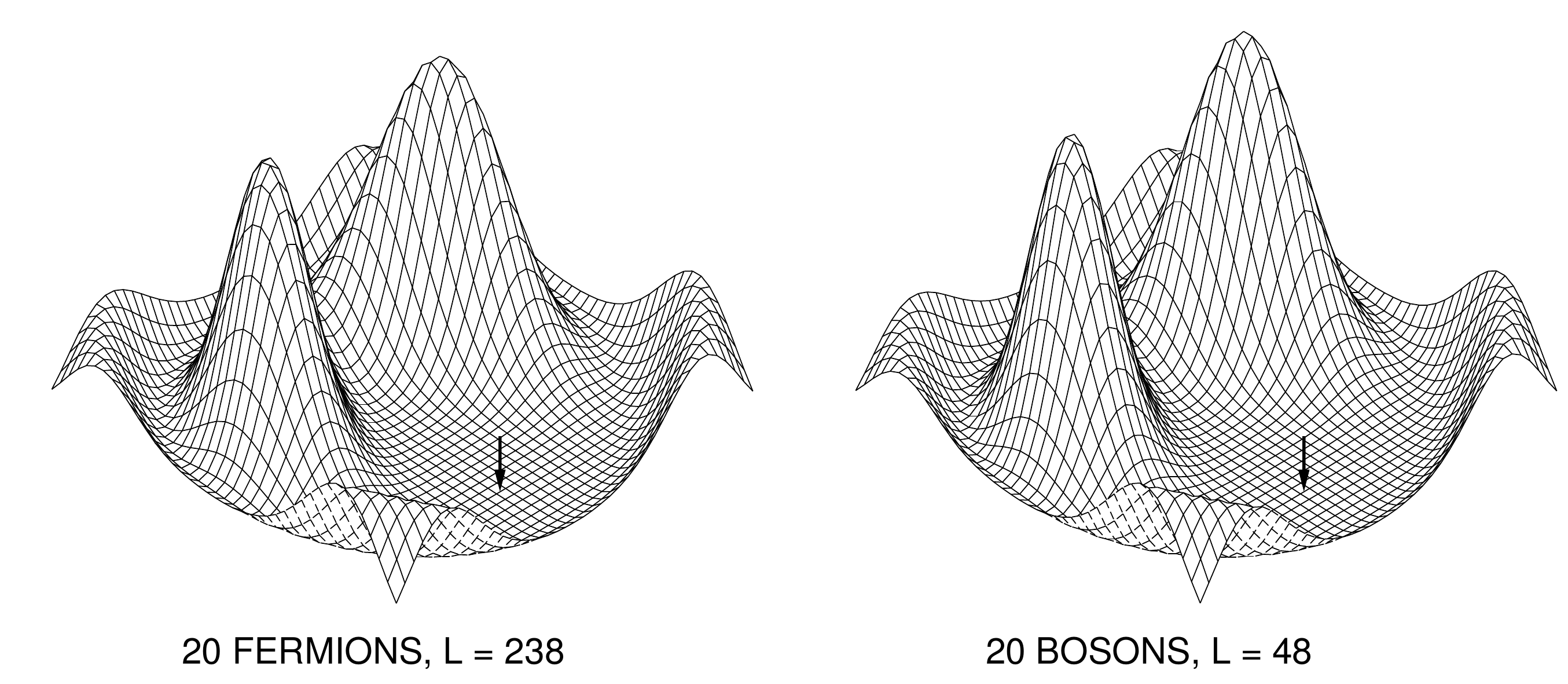}
\caption{Vortex-vortex correlation functions for three vortices
in a fermion and boson system with $N=20$ particles.  
(The difference in total angular momentum is due to the MDD in the 
fermion case, with $L_{\rm MDD}=N(N-1)/2$).
The boson wave function was first transformed to fermion Fock states, 
as described in the text. The arrows shows the site of the reference vortex.}
\label{ppbf}
\end{figure} 

For bosons, the localization of vortices is not as easily
seen in the pair correlations as for fermions -- mainly, 
because the bosonic occupancies in the Fock states make it difficult to
interpret the vortices directly as unoccupied states or holes in the MDD
(as in the fermion case),
as the occupation number is not limited. Nevertheless, for small 
particle numbers we can transform of the boson wave
function to its fermionic equivalent, by multiplying it with the
determinant $\prod (z_i-z_j)$. Figure~\ref{ppbf} shows the
vortex-vortex pair correlations determined in this way, 
for $N=20$ bosons with three vortices ($L=48$). For comparison, the 
corresponding fermion state (which in this case is not the ground state)
is shown as well ($L=328$). 
The correlation functions appear suprisingly similar. 

The above analysis showed very clearly that vortex localization 
occurs both in the bosonic and the fermionic case, and is mapped out very
directly by studying the corresponding correlation functions. 
The rotational spectra confirmed this observation. 

Figure~\ref{bv1} shows the energetically low-lying many-body energies 
for $N=20$ and 40 bosons, respectively. 
As for fermions, we observe the oscillatory behavior of the 
yrast line. We saw above that the period of 
the oscillations corresponds to the number of the first few  
vortices that localize on a ring. Note, that on the horizontal axis we
have now given the ratio $L/N$ in order to demonstrate that 
for bosons, the regions for different vortex numbers
only depend on $L/N$. This is not the case in fermion
systems~\cite{reimann2006b}, where for larger systems with $N\ge 14$, 
the vortices appear 
 closer of the surface of the MDD, leaving its center unaffected. 

Finally, we mention the possibility of vortex formation 
in boson and fermion systems where the particles have an internal
degree of freedom, like spin or pseudospin. In this case, 
the single-component vortex patterns are still observed, however, 
they are not any longer lowest-energy excitations.
This holds for fermions~\cite{koskinen2006} as well as for 
bosons.

Concluding our discussion of vortices in harmonically confined quantum 
systems that are set rotating, we should emphasize that the vortex
formation gives characteristic oscillations in the yrast 
spectrum~\cite{reimann2006b}.
The low-energy states of the rotational spectrum are determined
by the rigid rotation and vibrational states of Wigner molecules
of vortices~\cite{manninen2005}.
The vortex formation is similar for bosons and fermions and
it is nearly independent of the form of the 
repulsive interparticle interaction~\cite{toreblad2004,vorov2003}. 
 
\section{One-dimensional systems}

\subsection{1D harmonic oscillator}

Let us finally discuss interacting electrons confined by a
one-dimensional harmonic oscillator, as well as a quasi-one-dimensional
quantum ring. In an anisotropic oscillator,
$V_{\rm ext}=(1/2)m(\omega_x^2 x^2+\omega_y^2 y^2+\omega _z^2 z^2)$
choosing the frequencies  $\omega _y $ and $\omega _z$ of two spatial 
directions is so large that the 
particles occupy only the lowest state in the perpendicular direction, 
the system becomes effectively one-dimensional. 

\begin{figure}[h]
\includegraphics[width=1.1\columnwidth]{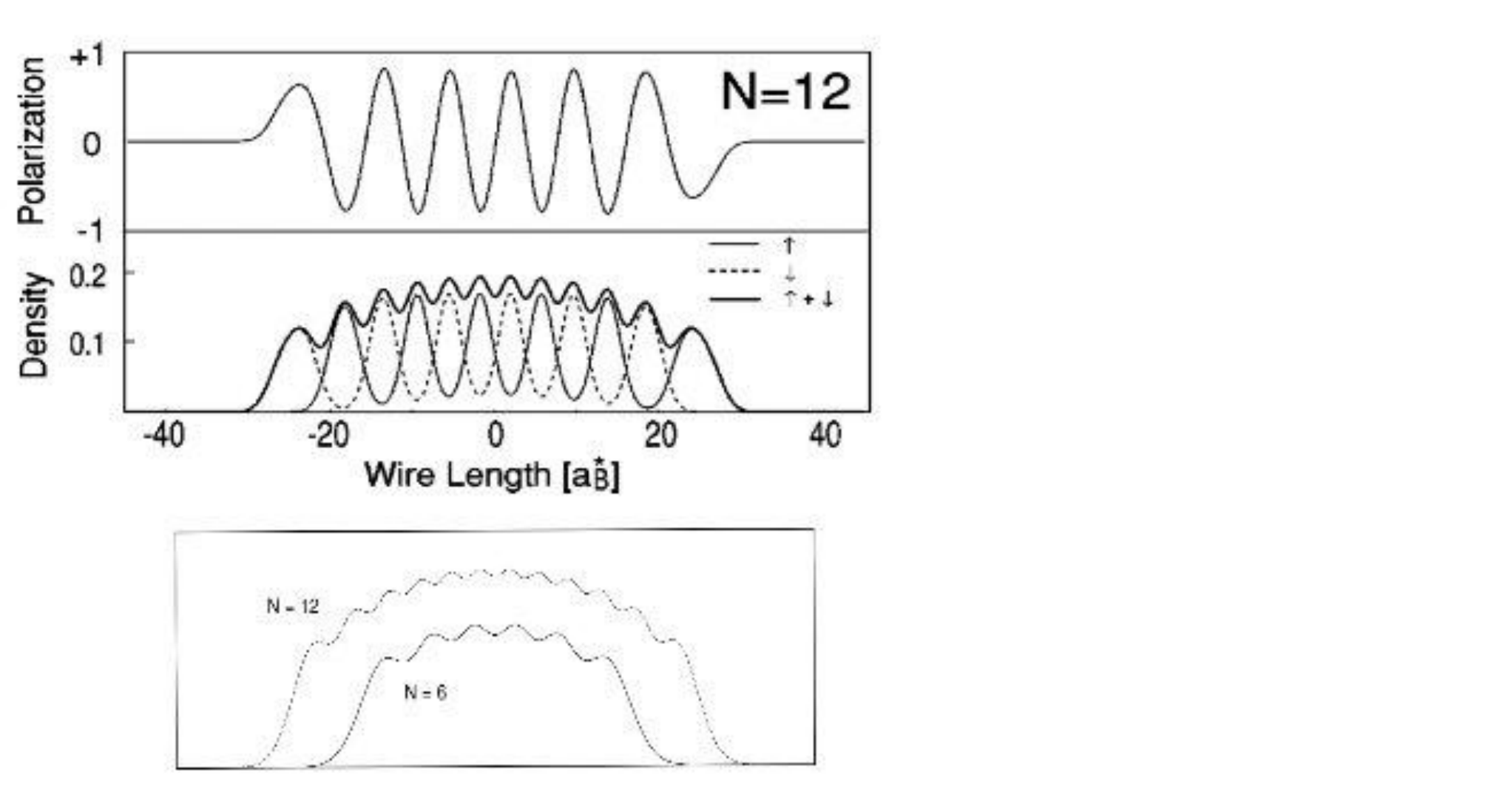}
\caption{Upper figure: Spin polarization, as well as spin- and total electron densities 
at the center of a quasi-1D wire with 12 electrons, calculated with
the LSDA. Lower figure: Electron densities for $N=6$ and $N=12$ noninteracting
spinless fermions in a 1D harmonic oscillator.}
\label{1dwire}
\end{figure} 

The 1D system of fermions is very different from the 2D and 3D cases.
The exchange interaction, or the Pauli exclusion principle, becomes
dominating. Since two electrons with the same spin can not be in the
same place, in 1D this means that electrons with the same spin can not 
pass each other. This enhances drastically the tendency to form a
spin density wave. In fact, an infinite 1D electron gas is unstable
against the so-called spin-Peierls transition: A static spin density
makes a spin-dependent mean-field potential (e.g. LSDA) with 
a wave length of $\pi/2k_F$ and consequently opens a gap at the Fermi level.
(Remember that the Fermi surface consist of only two points
in 1D).  
Figure~\ref{1dwire} shows the result of an LSDA calculation for 12
electrons in a quasi-1D harmonic potential, showing very clearly 
the resulting spin-density 
wave. The total density shows 12 maxima corresponding to 'localized'
electrons forming an anti-ferromagnetic chain. Even in 1D, the LSDA
can not properly localize the electrons. 

\begin{figure}[h]
\includegraphics[width=0.6\columnwidth]{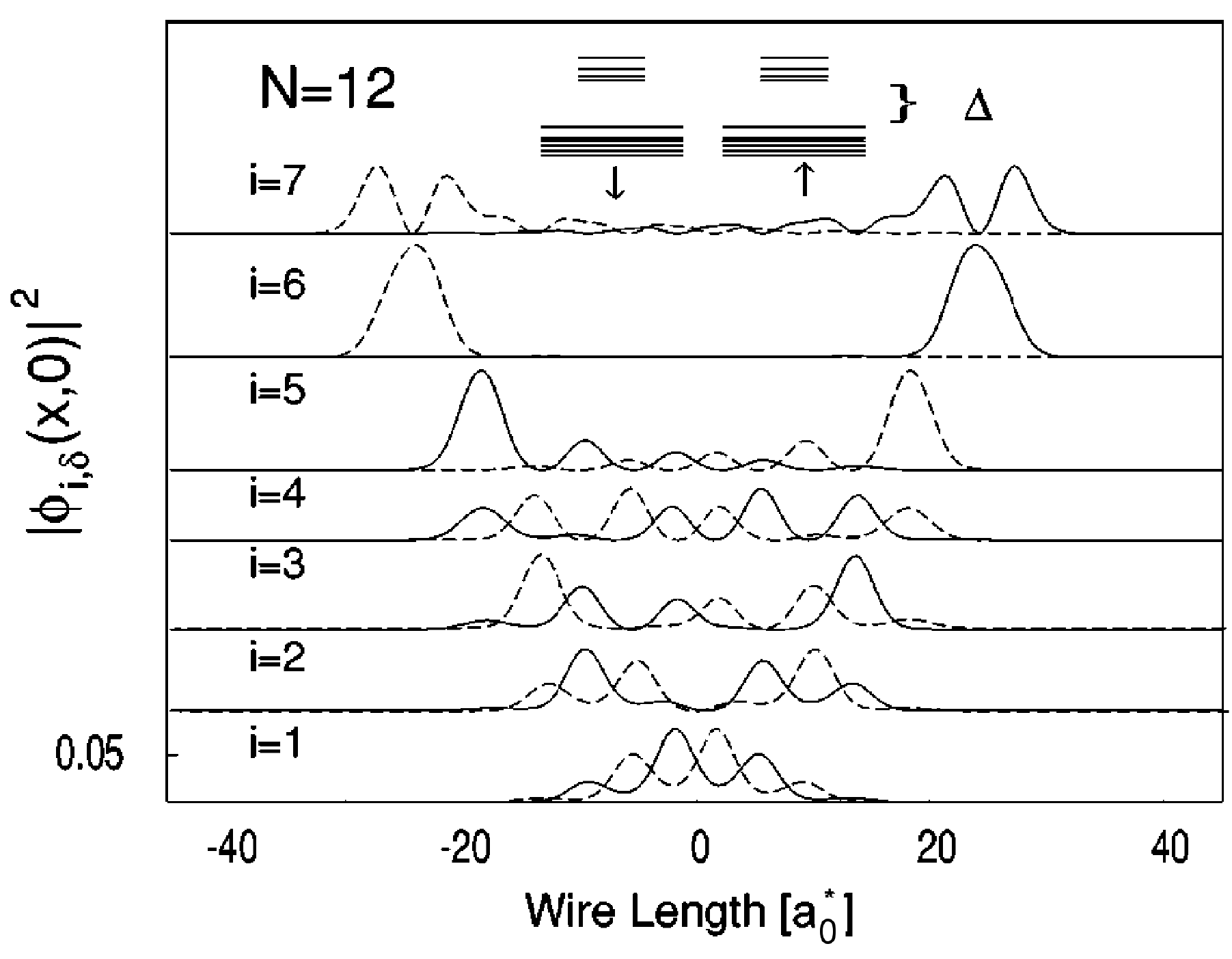}
\caption{Single-particle (Kohn-Sham) densities and energy
eigenvalues (inset) for a linear finite wire with
12 electrons. Note that the last occupied state ($i=6$) is localized 
at either end of the wire. The energy gap between occupied and
unoccupied states is denoted by $\Delta$ in the inset. From Ref.~\cite{reimann1998}.}
\label{1dhstates}
\end{figure} 

It is interesting to compare the self-consistent electron density 
to that of non-interacting electrons, shown also in Fig.~\ref{1dwire}.
The density for 12 {\it spinless} electrons is quite similar to the
LSDA result, while the density of 12 electrons with spin
has only 6 maxima since each single-particle state now occupies two  
electrons. However, the similarity of the LSDA density to that of
the noninteracting spinless electrons does not reach to the individual
single-particle wave functions. Figure~\ref{1dhstates} shows the
densities of the single-particle wave functions of the LSDA calculation.
Interestingly, the last occupied state $i=6$ is localized at the end
of the electron cloud. This {\it end state} is related to the
surface states in a metal surface. The existence of the periodic
potential which ends at the surface makes localized states
possible~\cite{zangwill1988}. In our 1D case the periodic potential is
provided by the spin-Peierls transition and the static spin-density
wave.

\subsection{Quantum rings}  
\label{sec:quantrings}

The observation of Aharonov-Bohm oscillations~\cite{aharonov1959}
and persistent currents~\cite{webb1985} have made quasi-1D
quantum rings a playground for simple theories. 
Indeed, the one-dimensionality as such gives a multitude
of interesting properties~\cite{kolomeisky1996, viefers2004}. 
Here we will only 
study the spectral properties of finite rings since
they are directly related to what we discussed earlier in 
connection to rotational states in a 2D harmonic confinement.

\begin{figure}[h]
\includegraphics[angle=-90,width=0.8\columnwidth]{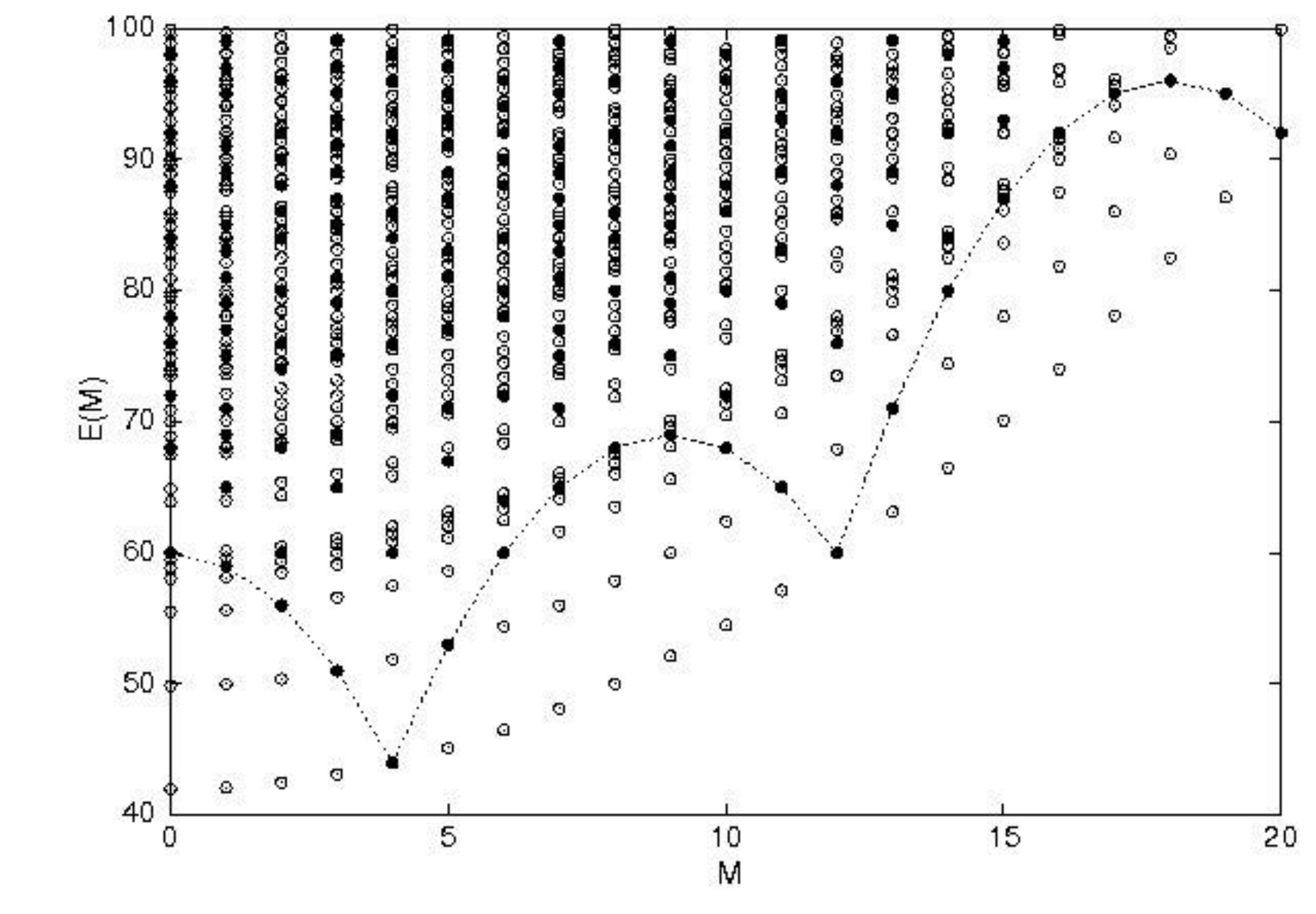}
\caption{Many-particle energy spectrum of 8 non-interacting
polarized electrons in a strictly one-dimensional quantum ring
(black dots) compared to the rotation-vibration spectrum of
classical particles interacting with $1/r^2$ interaction
(open circles). The dotted curve shows the yrast line of the 
polarized electrons.
From Ref.~\cite{viefers2004}.}
\label{ring8ni}
\end{figure} 

In the strictly 1D case, the single-particle eigenvalues are 
$\epsilon_l=\hbar^2l^2/2mR^2$, where $R$ is the radius of the ring 
and $l$ the angular momentum eigenvalue. The corresponding
single-particle states are $\psi(\phi)=\exp(il\phi)$.
The total angular momentum and energy for noninteracting
particles is 
\be
L=\sum_i^N l_i, \qquad E=\sum_i^N \epsilon_{l_i}.
\label{ni1d}
\ee
Let us first consider noninteracting polarized (spinless) fermions.
It is easy to determine their energy as a function of 
the total angular momentum using Eqs.~(\ref{ni1d}).
The results are shown for $N=8$ fermions in Fig.\ref{ring8ni}
as black dots.
The yrast line shows a period of eight, suggesting that the electrons
are localized in an octagon,
the downward cusps corresponding to purely rotational states
of the octagon. The black dots correspond to internal
vibrations of the Wigner molecule.
The fact that {\it noninteracting
polarized electrons form a Wigner molecule} is a special property 
of 1D. It can be shown that particles interacting with 
$1/r^2$ interaction in a 1D ring have the same energy spectrum
as noninteracting particles (or particles interacting
with an infinitely strong interaction of delta-function 
type)~\cite{viefers2004}.
Figure \ref{ring8ni}
shows (as open circles) also 
the classically determined energies
\be
E=E_{\rm rot}+E_{\rm vib}=\frac{\hbar^2L^2}{2NmR^2}
+\sum_\nu n_\nu\hbar\omega_\nu
\ee
for the $1/r^2$ interaction. 
Each vibrational level forms a rotational band. 
We can see that the spectrum of noninteracting polarized 
fermions (black dots) consists only of points at the classical
energies. In fact, for electron with spin and infinite strong
delta-function interaction 
($v(r)=A\delta(r)$, where $A\rightarrow\infty$) one obtains all the 
classical points (open circles).

The reason why noninteracting spinless electrons 
localize and have vibrational modes simply follows from the 
fact that the electrons can not pass each other. If the 
electron-electron distance is $d$, each electron 
is then localized between its neighbors in a region $2d$.
Its kinetic energy will then be proportional to $1/d^2$. This  
effectively leads to a $1/r^2$ interaction between the
electrons.

Interacting electrons in 1D systems have been extensively
studied using the Hubbard model (for reviews see
\cite{kolomeisky1996,viefers2004}). The energy spectrum can be solved
exactly using the Bethe Ansatz~\cite{lieb1968}. There, several 
analytic results exist. For a half-filled Hubbard band
(with one electron per site) it is rather easy to show that the large $U$-limit
the Hubbard model becomes an anti-ferromagnetic Heisenberg model. 
However, the Heisenberg model seems to be a good approximation
also for small filling~\cite{Yu1992,viefers2004}. This is important,
since the low-filling limit of the Hubbard model approaches to
free electrons with delta interaction (this is the same
as the tight-binding model approaching the free
electron model at the bottom of the band~\cite{manninen1991}).
Thus, also free electrons with spin localize in 
anti-ferromagnetic order as long as they have strong enough repulsive
interactions between them. 

\begin{figure}[h]
\includegraphics[width=0.6\columnwidth]{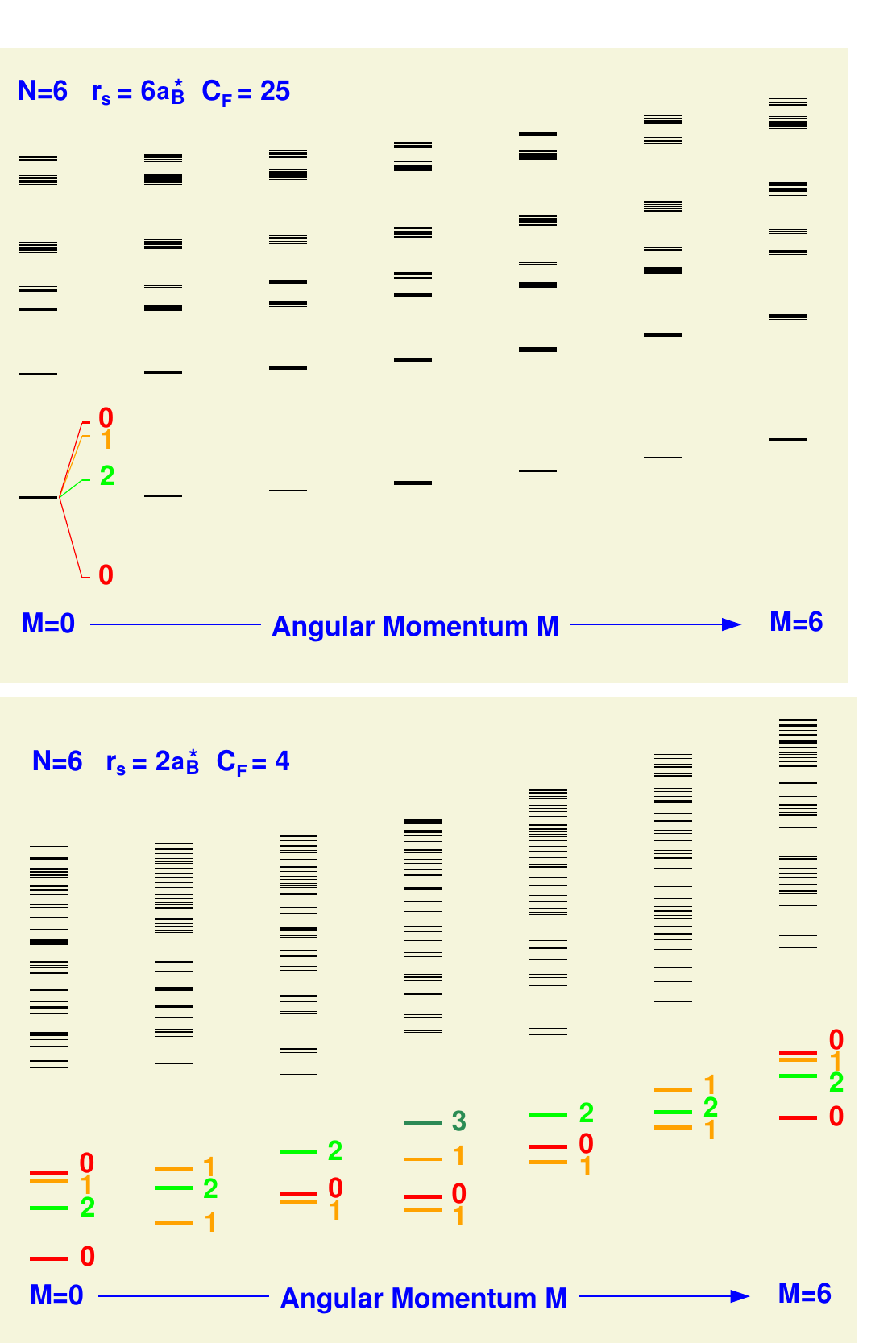}
\caption{Energy spectra for two quasi-one-dimensional
continuum rings with six electrons (in zero magnetic field).
The upper panel is for a narrow ring and it shows several
vibrational bands. The lower panel is for a wider ring 
which shows stronger separation of energy levels corresponding
to different spin states (shown as numbers next to the energy levels). 
Note that also the narrow ring has the same spin-ordering of the 
nearly degenerate state as expanded for the lowest $L=0$ state.
}
\label{ring}
\end{figure}

Koskinen {\it et al.}~\cite{koskinen2001} performed exact
diagonalization calculations for electrons confined in a quasi-1D
ring described with the external 2D potential
$V_{\rm ext}=m\omega_0^2(r-r_0)^2/2$, where $r=(x,y)$.
The rotational spectrum for six particles is shown in
Fig.~\ref{ring} for two different values of the narrowness of
the ring. The upper panel corresponds to a very narrow ring.
In this case, the different vibrational bands are clearly separated
and correspond quantitatively to the energies determined by solving the 
vibrational frequencies of the classical linear chain of electrons  
on the  ring.
The lower panel shows the result obtained for a wider, less one-dimensional 
ring. In this case,  
only the vibrational ground state is clearly separated, with the 
different spin-states separating  in energy.
With high accuracy, these  
different spin-states correspond  to those 
of an antiferromagnetic Heisenberg model for 
six  electrons on a ring~\cite{koskinen2001}. 

In narrow quantum rings the rotational spectrum is very robust.
It is insensitive  to the interparticle
interaction or the specific model for the confinement. Even the discrete
Hubbard model gives similar results as the continuum 
approaches~\cite{viefers2004}.
However, this demonstrates once more that the most clear indication of  
Wigner molecules in the ground states of high-symmetry systems
can be  obtained by analyzing the rotation-vibration spectrum.

\section{Concluding remarks}

In this short review, we summarized some characteristic aspects of 
finite quantal systems, that have their origin in the quantized level
structure in -- and beyond -- the mesoscopic regime. 
We discussed shell structure and deformation, as well as
the ocurrence of Hund's rule in finite fermion systems, conjointly for
metallic clusters, quantum dots in semiconductor heterostructures, or cold
atoms in traps -- seemingly different, but nevertheless in many aspects 
rather similar quantum systems. Like for atoms, shell structure does not only 
determine the stability and chemical inertness of metallic clusters, but also 
determines the conductance of a small quantum dot -- both close to, 
and far away from equilibrium.   

The experimental realization of Bose-Einstein condensation in an atomic 
gas~\cite{bec,bec2,traprmp1,traprmp2} opened up a whole new research field on 
ultra-cold atoms and coherent matter. In a cloud of bosonic atoms that is set
rotating, vortices may form. 
We discussed the fact that this vortex formation 
is not unique for bosonic systems, but may occur in a very similar way
for (non-paired) fermions under rotation, showing many analogies to the 
physics of the quantum Hall effect. 
Extreme rotation causes strong correlations, and the system is formally
equivalent to charged particles in a strong magnetic field. 
We finally gave a short summary of the physics of a finite fermionic system 
in quasi one dimension. 

As a final remark, we wish to emphasize that the many analogies 
existing between
nanostructures such as quantum dots and quantum wires, and cold atom gases
will become more important in the future -- last but not least 
due to the fact that these systems can be built much more ``clean'', and 
thus more coherent, than their semiconductor counterparts. 
An example for the cross-fertilization between these different sub-fields
of physics, is the recently discussed possibility of  
of {\it van-der-Waals blockade}~\cite{capelle2007}, 
which  is expected to play a key role in
transport experiments on confined cold atoms, and in {\em atomtronic} 
devices~\cite{atomtronics}.

\section{Acknowldegements}

This work was financially supported by the Swedish Research Council and
the Swedish Foundation for Strategic Research, as well as 
uthe European Community project ULTRA-1D 
(NMP4-CT-2003-505457). We thank J. Akola, M. Borgh, 
P. Singha Deo, H. H\"akkinen,
G. Kavoulakis, M. Koskinen, P. Lipas, B. Mottelson, P. Nikkarila,
M. Toreblad, S. Viefers, and Y. Yu 
for their collaboration on the 
subjects discussed in this review.

\end{document}